\documentclass[a4paper,11pt]{article}
\pdfoutput=1 

\usepackage{jheppub} 

\usepackage[T1]{fontenc} 
\usepackage{graphicx}
\usepackage{amsmath}
\usepackage{amssymb}
\usepackage{array}
\usepackage{slashed}
\usepackage{comment}
\usepackage{datetime}
\usepackage{soul}
\usepackage{epsfig}
\usepackage{epstopdf}
\usepackage{multirow}
\usepackage{pifont}

\usepackage{float}
\usepackage{bm}
\usepackage{bbm,multicol}

\usepackage{subfigure}
\usepackage{dsfont}

\usepackage{marginnote}

\newcommand{\met}{\ensuremath{{\not\mathrel{E}}_T}}

\def\h{h^0}
\def\H{H^0}
\def\A{A}
\newcommand{\cmark}{\ding{51}}%
\newcommand{\xmark}{\ding{55}}%

\title{Exotic Higgs Decay via Charged Higgs}
\author{Tong Li$^{\bf a}$}

\author{Shufang Su$^{\bf b}$}

\emailAdd{tong.li@monash.edu,  shufang@email.arizona.edu}
\affiliation{
$^{\bf a}$  ARC Centre of Excellence for Particle Physics at the Terascale,
 School of Physics and Astronomy, Monash University,
 Melbourne, Victoria 3800, Australia \\
$^{\bf b}$
Department of Physics, University of Arizona, P.O.Box 210081, Tucson, AZ 85721, USA\\
}

\abstract{
The most common search channel for heavy neutral Higgses in models with an extension of the Standard Model Higgs sector is $A/H^0\rightarrow \tau\tau$, which becomes ineffective when new decay modes of $A/H^0$ open. In this paper, we analyzed two such channels involving charged Higgses in the final states: $A/H^0 \rightarrow W^\pm H^\mp$ and $H^0 \rightarrow H^+H^-$. With the consequent decay of $H^\pm\rightarrow \tau\nu$, we found that the limits for $\sigma\times{\rm BR}(gg \rightarrow A/H^0 \rightarrow W^\pm H^\mp)\times {\rm BR}(H^\pm \rightarrow \tau \nu)$ vary from 30   to 10 fb for $m_{A/H^0}$ between 300 and 1000 GeV for 95\% C.L. exclusion, and about 80 to 30 fb for 5$\sigma$ discovery. For $H^+H^-$ mode, 95\% C.L. limits on $\sigma\times {\rm BR}(gg\to H^0\to H^+ H^-)\times {\rm BR}^2(H^\pm\to \tau\nu)$ vary from 9 to 4 fb for $m_{H^0}$ between 400 and 1000 GeV, while the 5$\sigma$ reach is about 20 to 10 fb. We further interpret the cross section limits in the Type II 2HDM parameter space. While $A\rightarrow W^\pm H^\mp$ offers great sensitivity in both $\sin(\beta-\alpha)$ versus $\tan\beta$ and $m_A$ versus $\tan\beta$ parameter space, $H^0\rightarrow H^+ H^-$ can cover most of the parameter space for $H^0$. Reach in $H^0\rightarrow W^\pm H^\mp$ is more limited, especially for $m_{H^0}>2 m_{H^\pm}$.   It is, however, complementary to $H^0\rightarrow H^+ H^-$ when ${\rm BR}(H^0\rightarrow H^+ H^-)$ is accidentally suppressed.
}

\begin{document}

\maketitle
\flushbottom
\newpage

\section{Introduction}
\label{sec:intro}

The discovery of the SM-like Higgs boson at the LHC is the greatest triumph in particle physics~\cite{Aad:2012tfa, ATLAS:2013sla, Chatrchyan:2012ufa,CMS:yva}. The stabilization of the observed Higgs mass of 126 GeV, however, provides strong motivation of physics beyond the Standard Model (SM). In addition, there are puzzles facing particle physics which cannot be explained in the SM, for example, the particle candidate for dark matter and the generation of neutrino mass. Solutions to those problems typically lead to models with an extended Higgs sector. Well known examples include the Minimal Supersymmetric Standard Model (MSSM)~\cite{Nilles:1983ge,Haber:1984rc,Barbieri:1987xf}, Next-to-Minimal Supersymmetric Standard Model (NMSSM)~\cite{Ellis:1988er,Drees:1988fc}, and Two Higgs Doublet Models (2HDM)~\cite{Branco:2011iw,type1,hallwise,type2}. In addition to a SM-like Higgs boson in these models, the low energy spectrum typically includes extra CP-even Higgses, CP-odd Higgses, as well as charged ones.

The discovery of beyond the SM Higgses is an unambiguous evidence for new physics beyond the SM.  The search for those extra Higgses, however, is typically   challenging. For the extra neutral Higgses at the LHC, most of the current searches focus on the conventional Higgs search channels of $WW$, $ZZ$, $\gamma\gamma$, $\tau\tau$ and $bb$~\cite{Aad:2014vgg,CMS-tautau,CMS-bb,atlas-WW2HDM,CMS-HZ,TheATLAScollaboration:2013wia,CMS_taunu}.  The production of the extra Higgses is typically suppressed compared to the SM Higgs, either due to their larger masses or their suppressed couplings to the SM particles.  The decays of beyond the SM Higgses to the $WW$ and $ZZ$ is absent for the CP-odd Higgs, and could be highly suppressed for the non-SM like CP-even Higgses.  The $\tau\tau$ or $bb$ decay modes suffer from either suppressed signal or large SM backgrounds, therefore are only relevant for regions of the parameter space with an enhanced $bb$ or $\tau\tau$ coupling. The search for the charged Higgs is even more difficult. For $m_{H^\pm}>m_t$, the cross section for the dominant production channel of $tbH^\pm$ is typically small.  The dominant decay mode $H^\pm\to tb$ is hard to identify given the large $tt$ and $ttbb$ backgrounds~\cite{Dev:2014yca}, while the subdominant decay of $H^\pm \to \tau\nu$ has suppressed branching fraction.  In the MSSM, even at the final stage of the LHC running, there is a wedge region in $m_A$ versus $\tan\beta$ plane with $\tan\beta\sim 10$, $m_{A}\gtrsim 300$ GeV in which only the SM-like Higgs can be observed at the LHC~\cite{Dawson:2013bba}.    Similarly, the reach for beyond the SM Higgses are limited in models with an extended Higgs sector.

In addition to the decays to the SM particles, beyond the SM Higgses can decay via ``exotic'' modes, i.e., heavier Higgs decays to two light Higgses, or one light Higgs with one SM gauge boson.  Examples include\footnote{Note that we use $\h$ and $\H$ to refer to the lighter or the heavier CP-even Higgs for models with two CP-even Higgs bosons.  When there is no need to specify, we use $H$ and $A$ to refer to the CP-even and CP-odd Higgses respectively.} $H\rightarrow AA$, $H \rightarrow H^+ H^-$, $H\rightarrow AZ$, $H\rightarrow W^\pm H^\mp$, $A\rightarrow HZ$, $A\rightarrow W^\pm H^\mp$, and $H^\pm \rightarrow AW, HW$ etc.  These channels typically dominate once they are kinematically open.  The current limits on the searches for beyond the SM Higgses are therefore weakened, given the suppressed decay branching fractions into SM final states.  Furthermore, these additional decay modes could provide new search channels for the beyond the SM Higgses, therefore complementary to the conventional search channels.

The study of the exotic Higgs decay modes have caught quite some attention recently.  Refs.~\cite{Coleppa:2013xfa,Coleppa:2014hxa,Brownson:2013lka,Dorsch:2014qja,Chen:2014dma, Chen:2013emb} studied $H\rightarrow AZ$, $A \rightarrow H Z$ decays in $bb\ell\ell$, $\tau\tau\ell\ell$, and $ZZZ$ final states.   Ref.~\cite{Coleppa:2014cca,Enberg:2014pua}  studied $H^\pm \rightarrow AW, HW$ decay from $tbH^\pm$ production for heavier charged Higgs.  The study of light charged Higgs produced in top quark decay from $t\bar{t}$ pair production or $tj$ single top production can be found in Ref.~\cite{AW_light}.  Earlier work for Higgs exotic decay can be found in Refs.~\cite{Maitra:2014qea,Basso:2012st,Dermisek:2013cxa,ATLAS_EARLY_STUDY,Assamagan:2002ne}.  It was found that those exotic Higgs decay modes could be complementary    to the conventional Higgs search modes, in particular, in regions of small $\tan\beta$. Thus they offer alternative Higgs search channels in regions where the conventional search channels are ineffective.  In particular, searches based on $A\rightarrow  Z \h$ and $\H \rightarrow \h \h$ has been performed at the ATLAS and CMS experiments~\cite{Aad:2015wra,CMS-HZ,CMS:2014yra}.

In this paper, we analyze two other exotic decay modes for the neutral Higgs that involve a light charged Higgs in the final states:
$A/H \rightarrow W^\pm H^\mp$ and $H \rightarrow H^+H^-$, with the subsequent decay of $H^\pm \rightarrow \tau \nu$. The SM backgrounds typically involve processes with $W\rightarrow \tau\nu$.  Due to the difference in the structure of Yukawa coupling and gauge coupling, and the spin correlation in tau decay, the charged decay product of tau lepton in the signal typically has a harder spectrum comparing to that of the backgrounds, which can be used to separate the signal from the SM backgrounds.

Note that there are strong flavor constraints on the mass of a light charged Higgs~\cite{Hpm_flavor}, in particular, $b\rightarrow s \gamma$~\cite{,Misiak:2015xwa}.    In addition, precision measurements require the charged Higgs to be nearly degenerate with one of the neutral Higgses~\cite{EW}. Those indirect    constraints, however, are typically model dependent and could be relaxed when there are contributions from other sectors in the model~\cite{Han:2013mga}.    In this paper, we focused on the collider aspect of beyond the SM Higgses and considered the light charged Higgs in the decay chain as long as it satisfies the current direct collider search limits.

The paper is organized as follows.  In Sec.~\ref{sec:2HDM}, we give a brief overview of the 2HDM and parameter regions where $A/H \rightarrow W^\pm H^\mp$ and $H \rightarrow H^+H^-$ can be important.  In Sec.~\ref{sec:limits}, we present the current collider limits on the neutral Higgses as well as the charged Higgs.  In Sec.~\ref{sec:analysis}, we present the details of our collider analyses.  In Sec.~\ref{sec:implication}, we study the implication of the cross section limits on the parameter space of the Type II 2HDM models.  Finally in Sec.~\ref{sec:conclusions} we summarize our main results and conclude.

\section{2HDM and Exotic Higgs Decay }
\label{sec:2HDM}

There are two ${\rm SU(2)}_L$ Higgs doublets in the 2HDM:
\begin{equation}
H_i=\left(
      \begin{array}{c}
        h_i^+ \\
        (v_i+h_i+iP_i)/\sqrt{2}
      \end{array}
    \right),  \ i=1,2.
\end{equation}
The neutral component of each Higgs doublet obtains a vacuum expectation value (vev) $v_1$ and $v_2$ with
$v=\sqrt{v_1^2+v_2^2}=246$ GeV after electroweak symmetry breaking.  Three degrees of freedom are eaten by the SM $W$ and $Z$ boson, with the remaining Higgses being two CP even Higgses $h^0$ and $H^0$, one CP odd Higgs $A$ and a pair of charged Higgses $H^\pm$:
\begin{equation}
  \left(
      \begin{array}{c}
        H^0 \\
        h^0 \\
      \end{array}
    \right)
= \left(
  \begin{array}{cc}
    \cos\alpha & \sin\alpha \\
   - \sin\alpha & \cos\alpha \\
  \end{array}
\right) \left(
      \begin{array}{c}
        h_1 \\
        h_2 \\
      \end{array}
    \right),  \ \ \
  \begin{array}{cc}
    A = & -\sin\beta \ P_1 + \cos\beta \ P_2 \\
    H^\pm = & -\sin\beta \ h_1^\pm + \cos\beta \ h_2^\pm
    \end{array}.
 \end{equation}

The most general Higgs potential has eight free parameters with the assumption of a discrete ${\cal Z}_2$ symmetry that can only be softly broken.  Two of these parameters are replaced by $v$ and $\tan\beta=v_2/v_1$, with the remaining five can be chosen as the CP-even Higgs mixing angle $\alpha$, physical Higgs masses $(m_{h^0}, m_{H^0}, m_A, m_{H^\pm})$ and the ${\cal Z}_2$ soft breaking parameter $m_{12}^2$.

The couplings that are relevant for $H^0/A \rightarrow H^\pm W^\mp$ are
\begin{equation}
W^\mp H^\pm H^0:  {g\over 2}\sin(\beta-\alpha)(p_{H^\pm}-p_{H^0})_\mu, \ \ \ W^\mp H^\pm A: {g\over 2}(p_{H^\pm}-p_{A})_\mu,\\
\end{equation}
with $g$ being the SM ${\rm SU}(2)_L$ coupling.   While the coupling of the charged Higgs to the heavy CP-even Higgs $H^0$ is proportional to $\sin(\beta-\alpha)$, the coupling of the charged Higgs to the CP-odd Higgs $A$ is independent of the mixing angle.  Note that if $H^0$ is non-SM like with $h^0$ being the observed 126 GeV SM-like Higgs, $|\sin(\beta-\alpha)| \sim 1$, which maximizes the $W^\mp H^\pm H^0$ coupling.   Given the strong limits on the light charged Higgs search either from the LEP~\cite{Hpm_LEP} or from the LHC \cite{TheATLAScollaboration:2013wia, CMS_taunu}, i.e. $m_{H^\pm}> 155$ GeV, we consider $m_{\H,A}>250$ GeV in the $h^0$-126 GeV case in our analyses.

The above couplings of gauge boson to a pair of Higgses are universal for different types of 2HDM.  The Higgs couplings to the fermions, however, is highly model dependent.  For the rest of paper, we will work in the framework of the Type II 2HDM, in which $H_1$ couples to the down-type quarks and leptons, while $H_2$ couples to the up-type quarks.  For a review of different types of 2HDM, please see Ref.~\cite{Branco:2011iw}.

\begin{figure}[h!]
\begin{center}
 \includegraphics[scale=1,width=7cm]{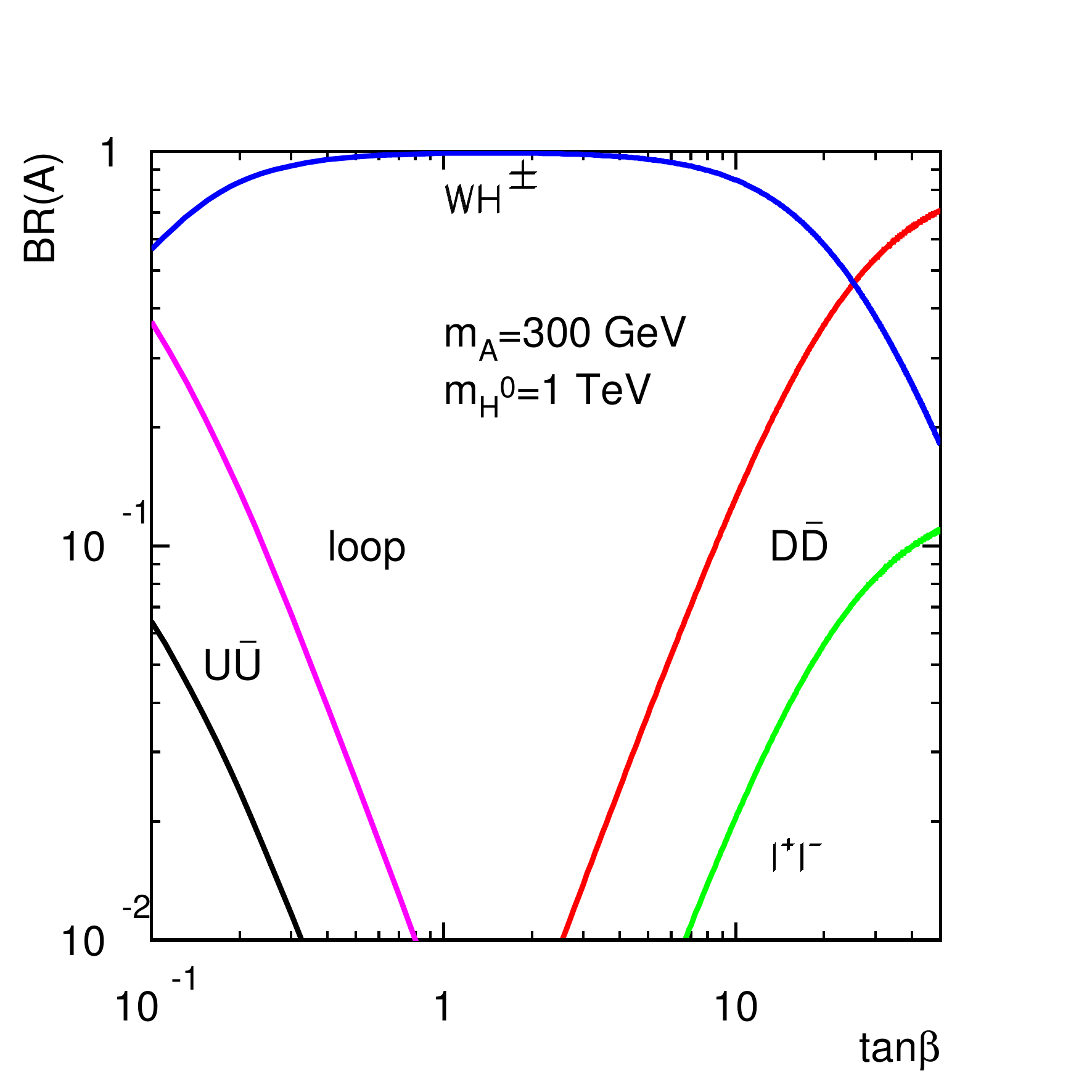}
 \includegraphics[scale=1,width=7cm]{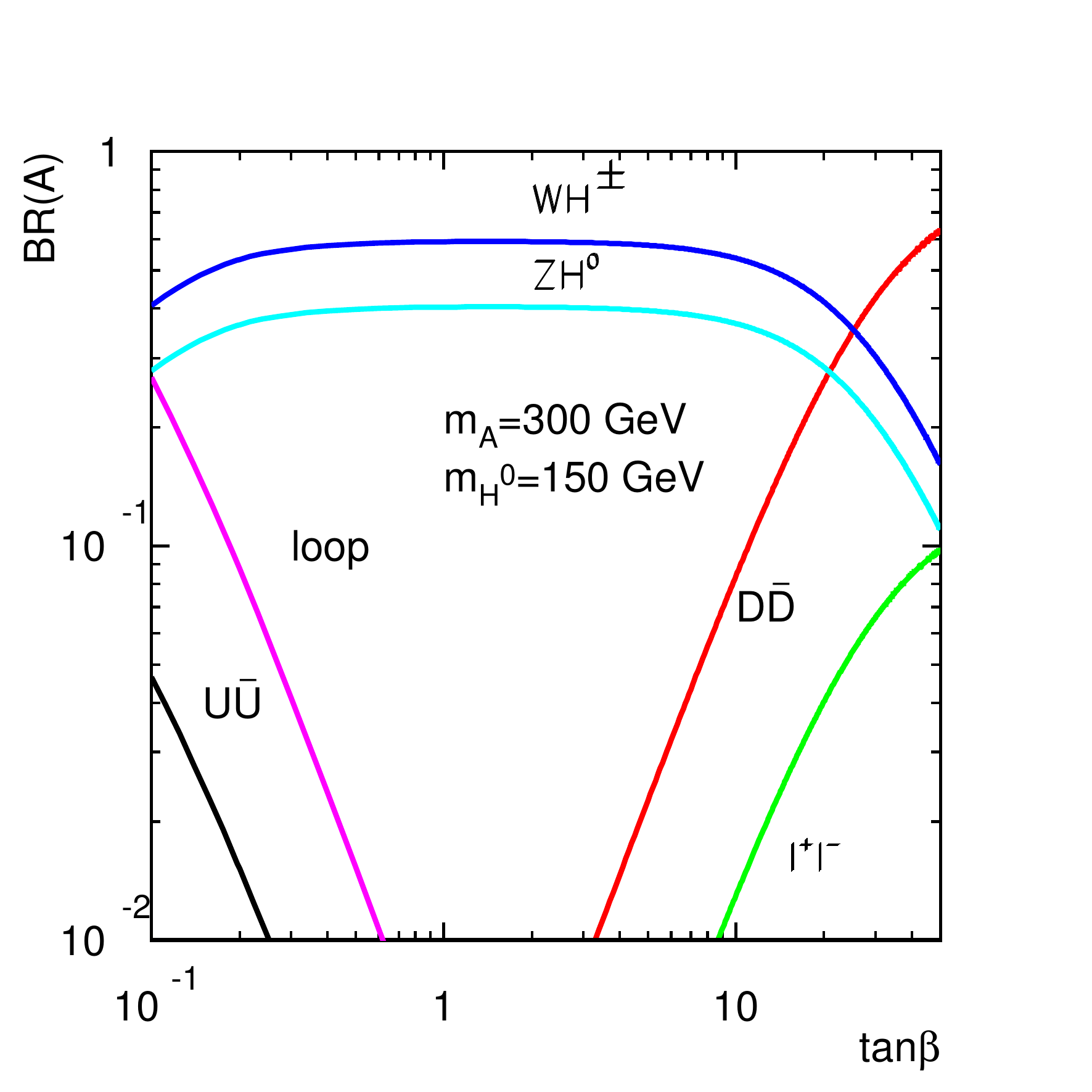}
 \end{center}
\caption{The branching fractions of $A$ as a function of $\tan\beta$ for $m_{A}=300$ GeV,  $m_{H^0}=$1 TeV (left panel) or 150 GeV (right panel). We have set $m_{H^\pm}=170$ GeV, $m_{h^0}=126$ GeV, $m_{12}^2=0$, and $\sin(\beta-\alpha)=1$.  ``loop'' indicates loop induced Higgs decays of $\gamma\gamma$, $Z\gamma$ and $gg$.
}
\label{fig:BRA0}
\end{figure}

Fig.~\ref{fig:BRA0} shows the branching fractions of $A$ as a function of $\tan\beta$. Once $A\rightarrow W^\pm H^\mp$ is kinematically open, in general, it quickly dominates over the usual fermionic decay mode $A\rightarrow bb,\tau\tau$ for $\tan\beta\lesssim 20$.  For large $\tan\beta$, branching fractions for $A\rightarrow bb,\ \tau\tau$ increase  due to  enhanced bottom and tau Yukawa couplings.    For small $\tan\beta$, the branching fraction for $A\rightarrow W^\pm H^\mp$ decreases slightly  since loop induced processes $A \rightarrow \gamma\gamma, Z\gamma$ rise due to enhanced top quark contributions.   The left panel of  Fig.~\ref{fig:BRA0} shows the branching fractions of  $A$ when  $H^0$ decouples.   Once $m_{A}>m_{H^0}+m_Z$, $A \rightarrow Z\H $ opens and gives a significant contribution in $A$ decay, as shown in the right panel of Fig.~\ref{fig:BRA0} for $m_{H^0}=150$ GeV.  $A \rightarrow H^\pm W^\mp$, however, still dominates, with branching fraction around 60\% in a large range of $\tan\beta$.

For the heavy CP-even Higgs $H^0$, in addition to the decay to $W^\pm H^\mp$ as the CP-odd Higgs $A$, it can also decay into a pair of charged Higgs once kinematically open: $H^0 \rightarrow H^+H^-$.  The relevant coupling receives contributions from all Higgs quartic couplings.  In general, it depends on $\tan\beta$, $\sin(\beta-\alpha)$, $m_{H^0}$, $m_{H^\pm}$ as well as $m_{12}^2$. Note that there is no $A H^+ H^-$ vertex because of the CP conservation assumption.

\begin{figure}[h!]
\begin{center}
\includegraphics[scale=1,width=7cm]{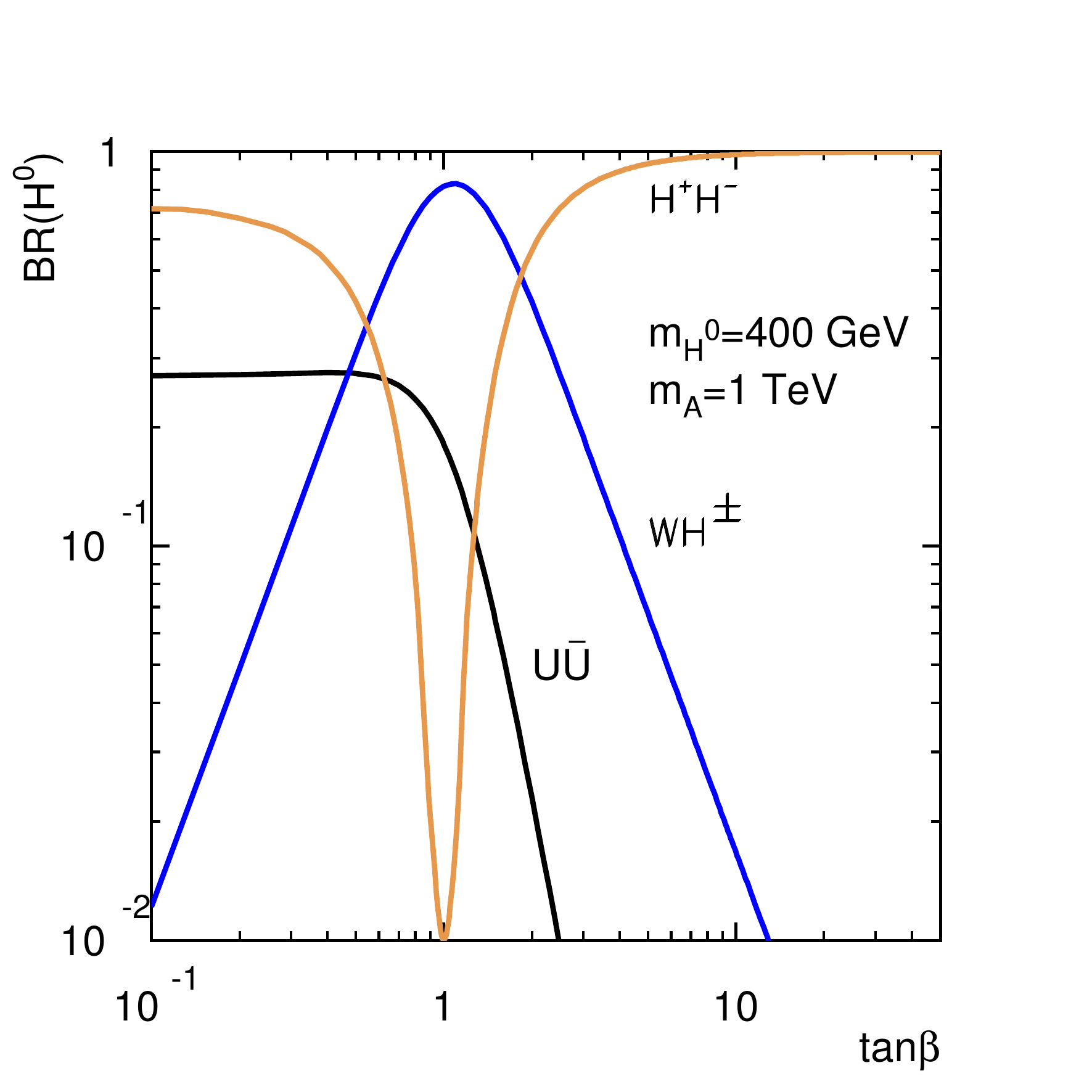}
\includegraphics[scale=1,width=7cm]{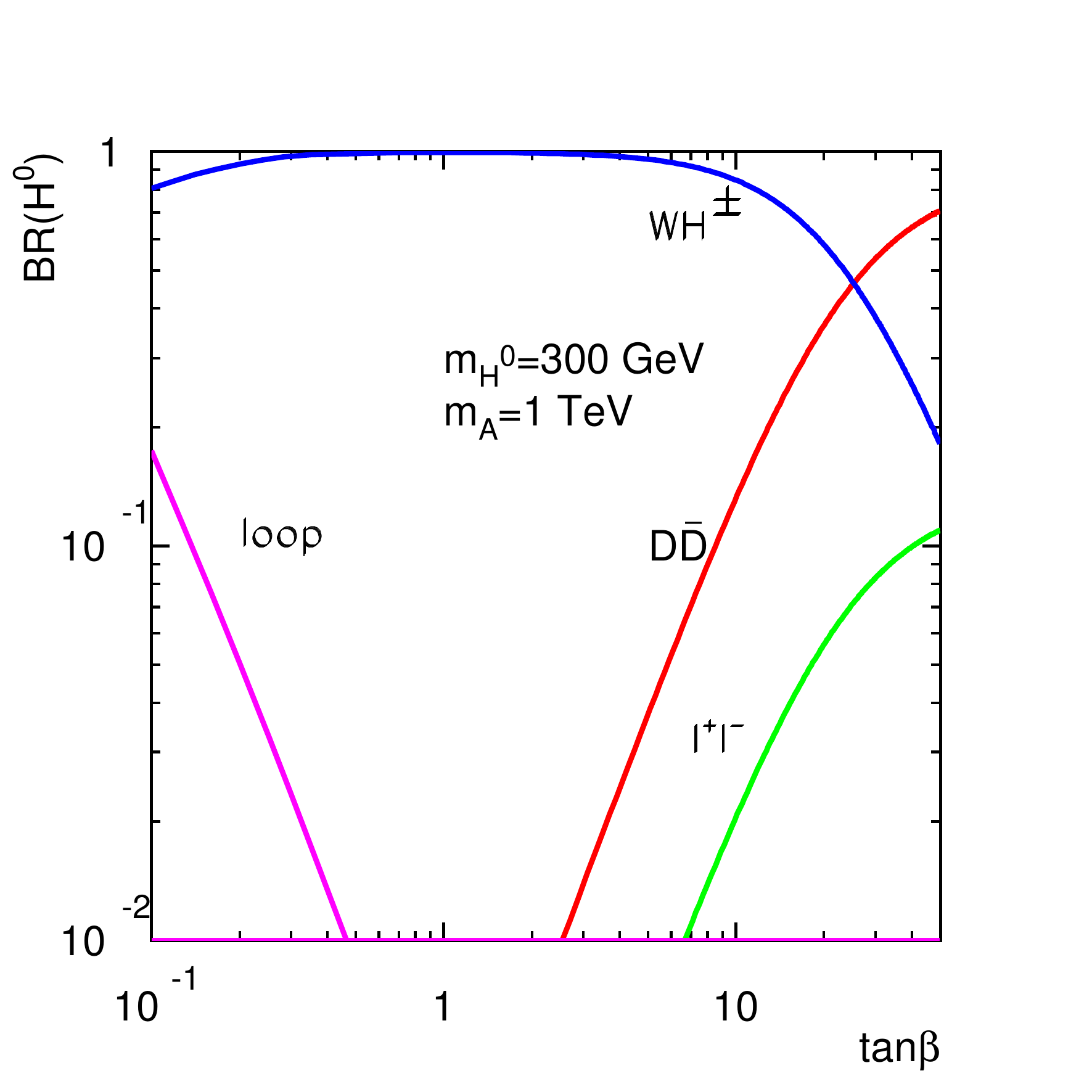}
 \end{center}
\caption{The branching fractions of $H^0$ as a function of $\tan\beta$ for $m_{H^0}=400$ (left panel) or 300 GeV (right panel). We have set $m_{H^\pm}=170$ GeV, $m_{h^0}=126$ GeV, $m_A=1$ TeV, $m_{12}^2=0$, and $\sin(\beta-\alpha)=1$.
}
\label{fig:BRH0}
\end{figure}

In Fig.~\ref{fig:BRH0},  the branching fractions of $H^0$ are shown as a function of $\tan\beta$ for $m_{H^0}=400$ GeV (left panel) and 300 GeV (right panel).   For $m_{H^0}=400$ GeV, $H^0 \rightarrow t\bar{t}$ opens up,  which dominates over $H^0 \rightarrow W^\pm H^\mp$ at small $\tan\beta$.  Once $H^0\rightarrow H^+H^-$ opens, it quickly dominates over other decay modes, unless there is accidental cancelation in $H^0H^+H^-$ couplings around $\tan\beta\sim 1$, as shown by the dip of the brown curve in the left panel of Fig.~\ref{fig:BRH0}.    With different values of  $m_{12}^2$, the suppression of  $H^0\rightarrow H^+H^-$ occurs at different values of $\tan\beta$.    When $H^0 \rightarrow H^+H^-$ is kinematically unaccessible, $H^0\rightarrow W^\pm H^\mp$ typically dominates over  the usual fermionic mode $H^0\rightarrow bb,\tau\tau$, except for large $\tan\beta$, as shown in the right panel of Fig.~\ref{fig:BRH0}.  For $m_{H^0}>m_{A}+m_Z$, $H^0$ could decay in addition to $AZ$, with decay branching fraction less than that of $H^0\rightarrow W^\pm H^\mp$.   For $m_{H^0}>2 m_{A}$, $H^0\rightarrow AA$ opens up, which could compete with $H^0\rightarrow H^+H^-$ and $H^0\rightarrow W^\pm H^\mp$.

Note that in the MSSM, in the decoupling region with $m_A\sim m_{H^0} \sim m_{H^\pm}$, $A/H^0\rightarrow W^\pm H^\mp$ and $H^0\rightarrow H^+H^-$ usually do not open due to the limited phase space.  These channels, however, could appear either in the MSSM with large loop corrections to the masses or in the NMSSM when extra singlet is introduced~\cite{Christensen:2013dra}.

\section{Current Collider Limits}
\label{sec:limits}

The main search mode for the non-SM like  neutral Higgses is through $\tau\tau$ channel, which has been performed at both the ATLAS and CMS experiments~\cite{Aad:2014vgg,CMS-tautau} with 7$+$8 TeV data set of about 20 ${\rm fb}^{-1}$ integrated luminosity.   For the gluon fusion production, the cross section limits of $gg\rightarrow \phi \rightarrow \tau\tau$ from ATLAS vary between 30 pb to 8 fb for $m_\phi$ between 100 to 1000 GeV.    Interpreting the cross section limits in the MSSM $m_h^{\rm{max}}$ scenario, a  sizable portion of the MSSM parameter space has been ruled out, extending from $\tan\beta\sim 10$ for $m_{\A} \sim$ 100 GeV, to $\tan\beta \approx 60$ for $m_{\A}$= 1000 GeV~\cite{Aad:2014vgg}. Limits from CMS are similar~\cite{CMS-tautau}.

\begin{figure}[h!]
\begin{center}
 \includegraphics[scale=1,width=7cm]{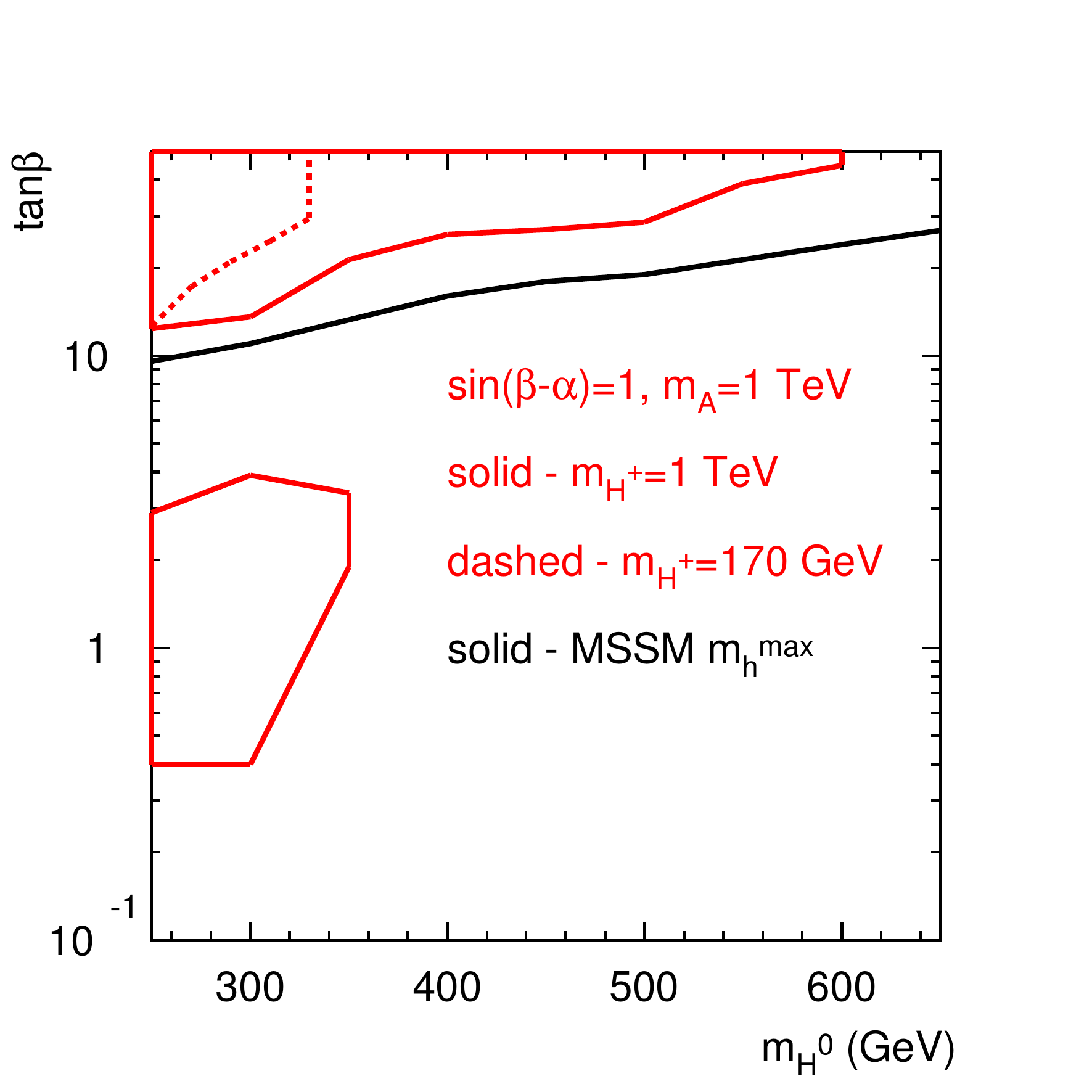}
\includegraphics[scale=1,width=7cm]{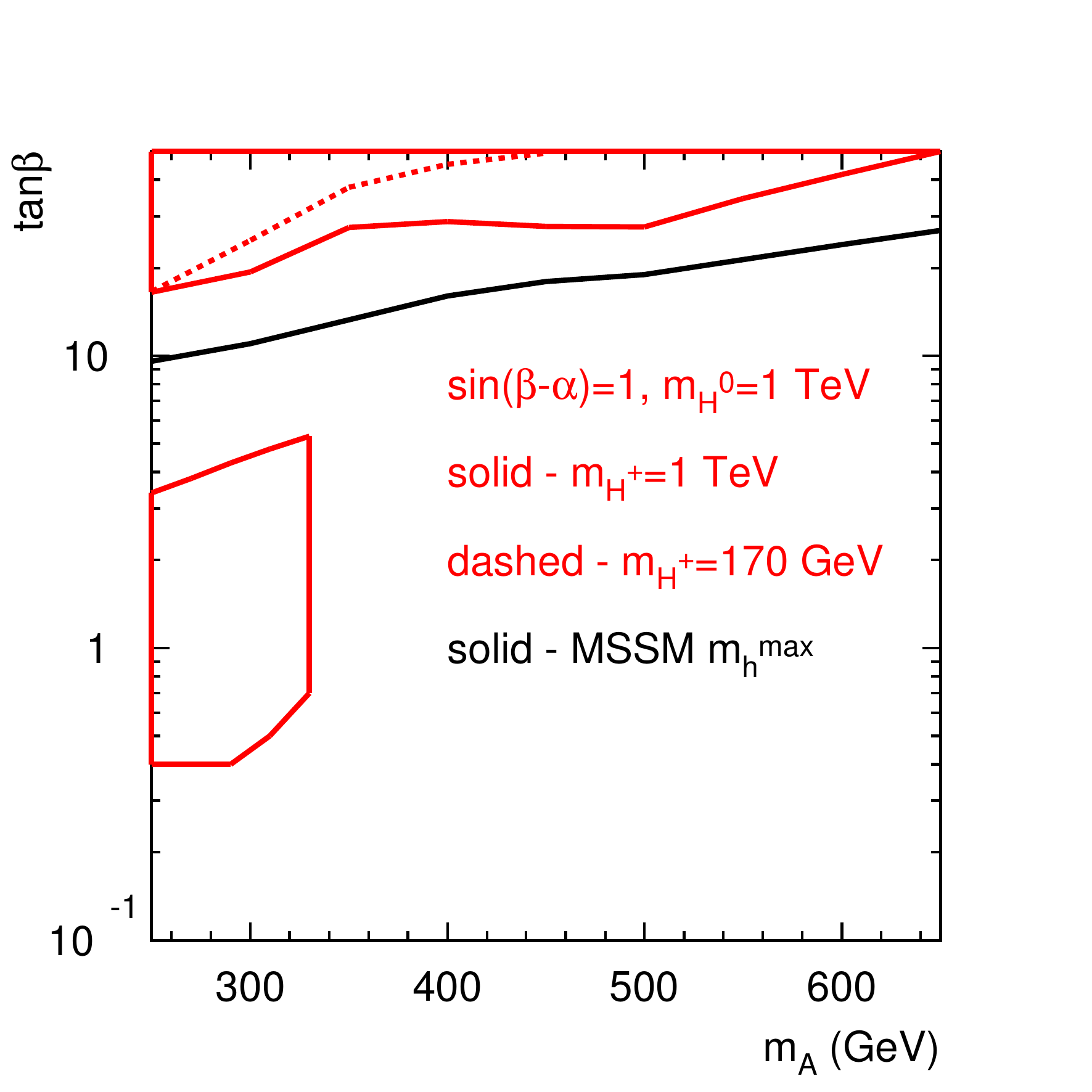}
 \end{center}
\caption{The excluded regions  for $H^0$ (left) and $A$ (right) in the Type II 2HDM  based on  $H^0/A\rightarrow\tau\tau$ searches at the 7$+$8 TeV LHC with  20 ${\rm fb}^{-1}$ integrated luminosity~\cite{Aad:2014vgg}.   The limits for the MSSM $m_h^{\rm{max}}$ scenario are  shown in  solid black curves, when both $A$ and $H^0$ contribute to the signal.   The solid red curves show the limits in the Type II 2HDM when only $H^0$ or $A$ contributes and $m_{H^\pm}=1$ TeV.  The dashed red curves show the relaxed limit for $m_{H^\pm}=170$ GeV when $A/H^0\rightarrow W^\pm H^\mp$ and $H^0\rightarrow H^+H^-$ are open.  Other parameter choices are $m_{h^0}=126$ GeV and $\sin(\beta-\alpha)=1$.
  }
\label{limit_A0H0}
\end{figure}

In  Fig.~\ref{limit_A0H0}, we recast the current 95\% C.L. limit   of $pp\to \phi \to \tau\tau$ in the $(m_{H^0},\tan\beta)$ (left panel) and $(m_A,\tan\beta)$ (right panel) planes of the Type II 2HDM.  The solid black curves correspond to the limits in the MSSM, when $m_A \approx m_{H^0}$ with both $A$ and $H^0$ contributing to the signal.  The solid red curves correspond to the limits in the type II 2HDM, when only contribution from $H^0$ or $A$ is included and other non-SM Higgses decouple.   The reach is considerably weaker: the current exclusion is about $\tan\beta\sim 12 $ at $m_{H^0}= 200$ GeV, and $\tan\beta \sim 50$ for $m_{H^0} = 600$ GeV, and similar for the CP-odd Higgs $A$.    Once $A/H^0\rightarrow W^\pm H^\mp$ and $H^0\rightarrow H^+H^-$ open for $m_{H^\pm}=170$ GeV, the limits are much more relaxed, as shown by the red dashed curves.    In particular, no limit on $m_{H^0}$ above $H^+H^-$ threshold can be derived given the strong suppression of $H^0\rightarrow \tau\tau$ branching fraction.

Searches with $bb$, $WW$, $h^0Z$ and $h^0h^0$ for the non-SM Higgses have also been performed at both the ATLAS and CMS~\cite{CMS-bb,atlas-WW2HDM,CMS-HZ,Aad:2015wra,CMS:2014yra}.  No evidence for a neutral non-SM like Higgs was found and the limits are considerably weaker than the $\tau\tau$ channel.

In our analyses, we consider $A/H^0\rightarrow W^\pm H^\mp$ and $H^0 \rightarrow H^+ H^-$, with a benchmark point of $m_{H^\pm}=170$ GeV.   Both ATLAS and CMS have searched for a charged Higgs in $H^\pm \rightarrow \tau\nu, cs$ mode~\cite{TheATLAScollaboration:2013wia,CMS_taunu}.  For the low mass region of $m_{H^\pm}<m_t$, the charged Higgs is produced via top decay, while for the high mass region of $m_{H^\pm}>m_t$, $tbH^\pm$ associated production is considered.   While most of the low mass region has been excluded for charged Higgs mass less than about 155 GeV,  there is no limit for 160 GeV $<m_{H^\pm}<m_t$ given the suppressed ${\rm BR}(t\rightarrow b H^\pm)$ near the threshold.  The reach for high mass region only extends down to $m_{H^\pm}>180$ GeV, due to the overwhelming SM $t\bar{t}$ background at the low mass region.

In addition, there are strong constraints on the non-SM Higgs sector from flavor constraints~\cite{Hpm_flavor} and precision measurements~\cite{EW}.  In particular, the latest analyses on ${\rm Br}(B\rightarrow X_s \gamma)$ with updated NNLO QCD predictions have constrained the charged Higgs to be heavier than 480 GeV at 95\% C.L.~\cite{Misiak:2015xwa}.  Precision observables, in particular, $S$ and $T$ oblique parameters, also impose correlations between the charged Higgs mass with the neutral ones: $m_{H^\pm} \sim m_{A}$ or $m_{H^\pm} \sim m_{H^0}$.  These limits, however, could be relaxed with additional contributions to the flavor or precision observables from other sectors in the new physics models~\cite{Han:2013mga}.  In this paper, since we focused on the collider aspect of the Higgs exotic decay, we did not apply those constraints explicitly.   We chose the mass spectrum of the non-SM Higgses to be characteristic of the exotic decay channels that we analyze.
One should, however, keep those potentially dangerous indirect constraints in mind when considering a specific new physics model with an extended Higgs sector.

\section{Collider Analyses}
\label{sec:analysis}
In this section we analyzed $gg\rightarrow A/\H  \rightarrow W^\pm H^\mp$ and $gg\rightarrow H^0 \rightarrow H^+ H^-$, with the subsequent decay of $H^\pm \rightarrow \tau \nu$. Due to the spin correlation in $\tau$ decay, the  charged product from tau  decay is typically harder for the ones from the signal process  with tau coming from $H^\pm$ decay comparing to the ones from the SM backgrounds with $\tau$ mostly from $W$ decay. Both signal and background processes are generated by MadGraph/MadEvent~\cite{MGME} and then passed to TAUOLA to simulate tau lepton decay~\cite{Tauola}. We present model independent limits on $\sigma\times{\rm BR}$ for both the 95\% C.L. exclusion as well as 5$\sigma$ discovery at the 14 TeV LHC with 300 ${\rm fb}^{-1}$  integrated luminosity.

\subsection{$gg \rightarrow A/\H \rightarrow W^\pm H^\mp$}
\label{sec:HpmWmp}

 We studied the gluon fusion production of $A/\H$, followed by $A/\H \rightarrow W^\pm H^\mp$ with $H^\pm \rightarrow \tau \nu$ and  $W \rightarrow \ell \nu$:
\begin{eqnarray}
&&gg\to A/\H\to W^\pm H^\mp  \to \ell^\pm\tau^\mp\nu\bar{\nu}, \ \ \ \ell=e,\mu.
\end{eqnarray}
The $\tau$ hadronic decays are adopted to take advantage of spin correlation for the final state hadrons.
We consider the two leading hadronic decay modes of tau lepton: $\tau^\pm \to \pi^\pm \nu_\tau$ and $\tau^\pm\to \rho^\pm \nu_\tau$ with ${\rm BR}(\tau^\pm\to \pi^\pm \nu_\tau)=0.11$ and ${\rm BR}(\tau^\pm\to \rho^\pm \nu_\tau)=0.25$.  For simplicity, we only display the results of the $\pi$ channel  below.   The contributions from both $\pi$ and $\rho$ channels are combined for the model independent $\sigma\times{\rm BR}$ limits.

The leading SM backgrounds are
\begin{eqnarray}
W^+W^-\to \ell^\pm  \tau^\mp  \nu  \bar{\nu}, \ W^+W^-\to \tau^+\tau^-\nu\bar{\nu} , \ ZZ\to \tau^+\tau^-\nu\bar{\nu}
\end{eqnarray}
with leading-order (LO) cross section $\sigma_1(WW)=3.64$ pb, $\sigma_2(WW)=0.91$ pb and $\sigma(ZZ)=0.3$ pb, including subsequent $W$ and $Z$ decay.  The latter two are followed by one tau decaying leptonic   and the other tau decaying hadronically.  The reducible backgrounds are
\begin{equation}
W^\pm Z\to \ell^+\ell^- \tau^\pm\nu,  \ell^\pm  \tau^+ \tau^- \nu
\end{equation}
with cross section $\sigma_1(WZ)=\sigma_2(WZ)=0.21$ pb including subsequent decay.   These processes have additional $e/\mu$ or $\tau$ lepton and can thus be reduced by vetoing the extra lepton. We apply the $K$-factors of 1.5, 1.3 and 1.7 to the channels $WW$, $ZZ$ and $WZ$, respectively~\cite{Dixon:1999di}.

We  select events with  one lepton and one hadronically decaying tau satisfying the basic cuts:
\begin{equation}
p_T(\ell)\geq 15 \ {\rm GeV}, \ |\eta(\ell)|<2.5; \ p_T(h_\tau)\geq 20 \ {\rm GeV}, \ |\eta(h_\tau)|<2.3;\ \Delta R_{\ell h_\tau} \geq 0.4,
 \label{basic1}
 \end{equation}
 where $h_\tau=\pi,\rho$.   We veto events with extra leptons or hadronically decaying taus satisfying
\begin{eqnarray}
{\rm veto:}\  p_T(\ell)> 7 \ {\rm GeV}, \ |\eta(\ell)|<3.5; \ p_T(h_\tau)> 10 \ {\rm GeV}, \ |\eta(h_\tau)|<4.9;
\label{veto1}
\end{eqnarray}
 To simulate the detector effects, we smear the hadronic/leptonic energy by a Gaussian distribution whose width is parameterized as~\cite{smearing}
\begin{eqnarray}
{\Delta E\over E}={a\over \sqrt{E/{\rm GeV}}} \oplus b, \ \ a_{had}=100\%,
\ b_{had}=5\%, \ a_\ell=5\%, \ b_\ell=0.55\%.
\end{eqnarray}

\begin{figure}[h!]
\begin{center}
\includegraphics[scale=1,width=7cm]{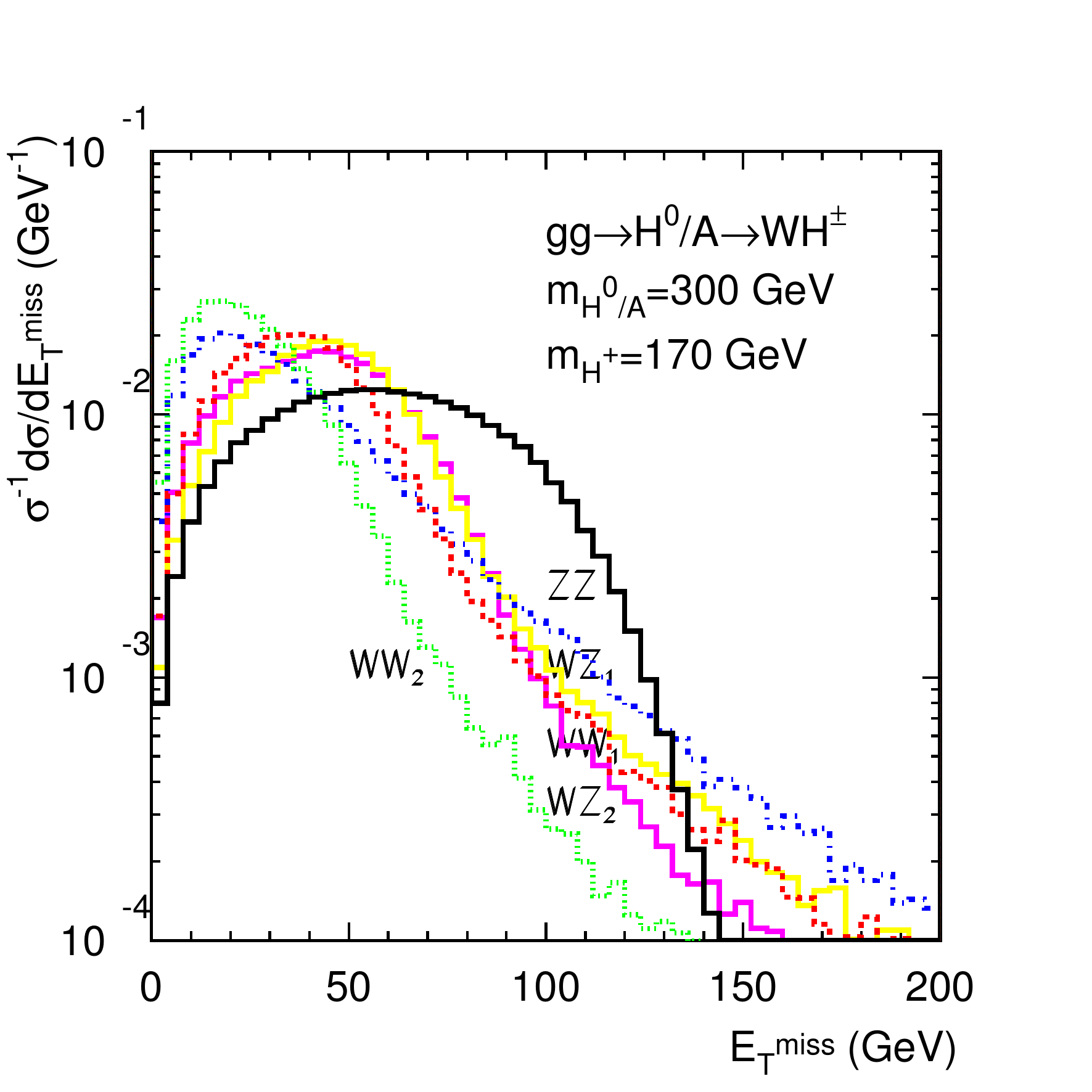}
\includegraphics[scale=1,width=7cm]{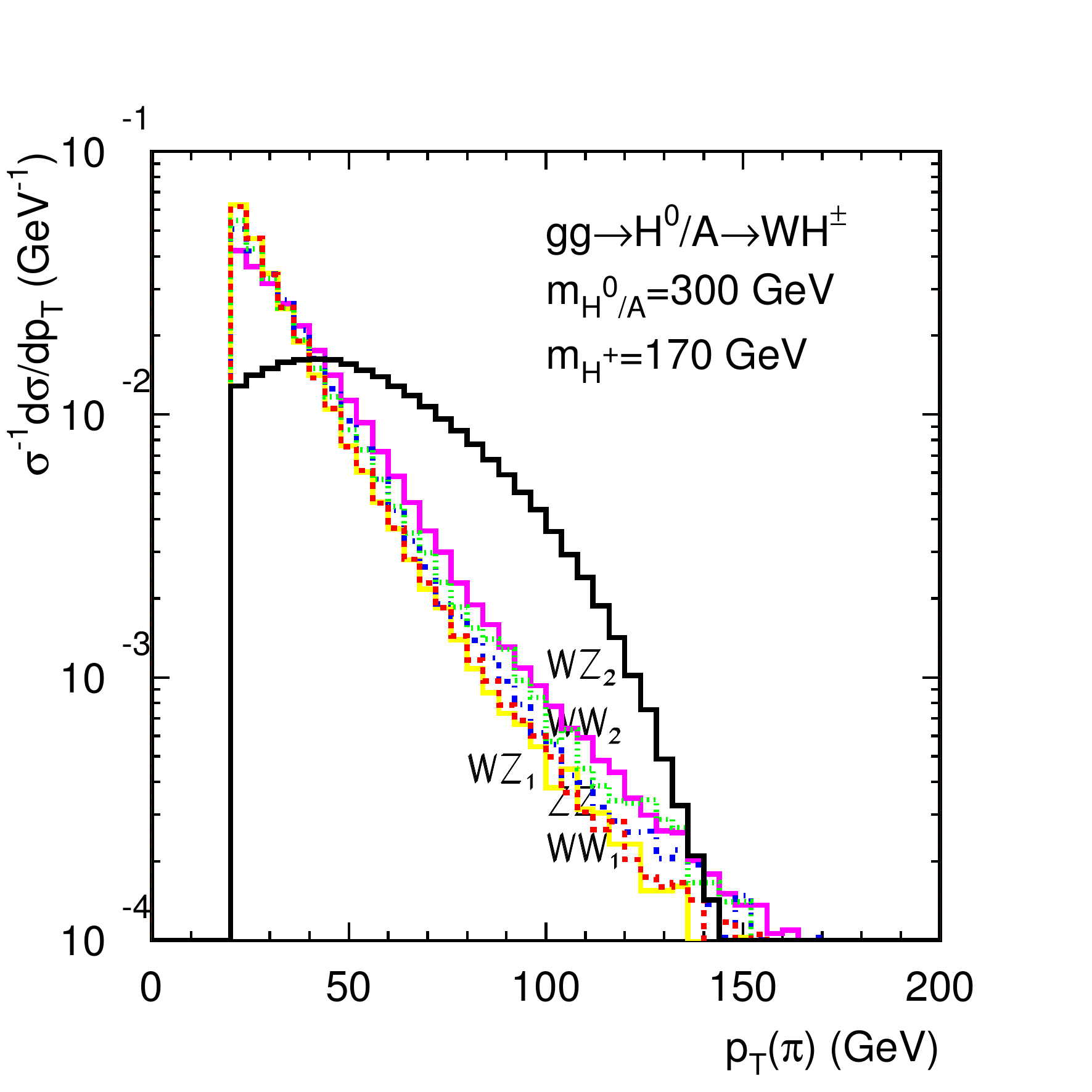}
\end{center}
\caption{The distributions of $p_T(\pi)$ (left panel)  and $\met$ (right panel) for signal $W^\mp H^\pm$ (black) and SM backgrounds after basic cuts in $\tau\rightarrow \pi \nu$ final states.  We assume $m_{H^0/A}=300$ GeV and $m_{H^\pm}=170$ GeV.
}
\label{events14}
\end{figure}

\begin{table}[h!]
\begin{center}
\resizebox{15cm}{!} {
\begin{tabular}{|c|c|c|c|c|c|c|c|c|c|}
  \hline
  Cuts & Signal & $(WW)_1$ & $(WW)_2$ & $ZZ$ & $(WZ)_1$ & $(WZ)_2$ &  $(WZ)_{2\ast}$  & $S/B$ & $S/\sqrt{B}$\\
  \hline
  $\sigma\times$BRs (fb) & 10 & 591 & 104 & 30 & 39 & 4.3 & 13.6 & 0.013 & 6.2\\
  \hline
 basic & 0.67 & 0.13 & 0.068 & 0.11 & 0.0065 & 0.056 & 0.0076 & - & - \\
 \hline
 $\met$ & 0.61 & 0.28 & 0.09 & 0.26 & 0.23 & 0.31 & 0.31 & - & -\\
 \hline
 $p_T(\pi)$ & 0.6 & 0.2 & 0.51 & 0.5 & 0.135 & 0.33 & 0.97 & - & -\\
 \hline
 \hline
  $\sigma\times$BRs (fb) & 2.45 & 4.2 & 0.33 & 0.43 & 0.0078 & 0.024 & 0.0079 & 0.49 & 19 \\
 \hline
\end{tabular}}
\end{center}
\caption{The cross sections (row 2 and 6)  and cut efficiencies (row 3$-$5)  of signal $gg\to A/\H \to W^\mp H^\pm\rightarrow \ell\tau\nu\nu$ and SM backgrounds after various cuts with $\tau^\pm\to \pi^\pm\nu$ at the 14 TeV LHC.  We assume a nominal signal cross section of 10 fb. For the background processes, the $K$-factors have been included.   $(WZ)_2$ and $(WZ)_{2*}$ refer to $WZ\rightarrow \ell^\pm  \tau^+ \tau^- \nu$ with both taus decay to pions for $(WZ)_2$  and one tau to pion and one tau to lepton for  $(WZ)_{2*}$.   The significance $S/\sqrt{B}$ is given for 300 ${\rm fb}^{-1}$ luminosity. We assume $m_{H^0/A}=300$ GeV and $m_{H^\pm}=170$ GeV.
  }
\label{cuts1}
\end{table}

With the above basic cuts and smearing, the distributions of $p_T(\pi)$ and $\met$  for the signal (black curve) and SM backgrounds are shown in Fig.~\ref{events14}.
We note that the signal has a harder $p_T(\pi)$ spectrum compared to the backgrounds.
This is a well-known result of spin correlation in the $\tau$ decay. For the $H^{+}$ signal, the left-handed $\tau^{+}$ decays to a right-handed $\bar\nu_{\tau}$, causing the $\pi^{+}$ to preferentially move along the $\tau^{+}$ momentum direction~\cite{KH}. In contrast, the $\tau^{+}$ coming from a $W^+$ decay is right-handed which has the opposite effect on the $\pi^+$. The distribution for $p_T(\rho)$ is similar for $\tau^\pm\rightarrow \rho^\pm \nu$.  Signal also has $\met$ distribution peaked at higher value. We thus tighten the selection cuts by imposing
 \begin{eqnarray}
 \met>50 \ {\rm GeV}, \ \ p_T(\pi, \rho)>50 \ {\rm GeV}.
\label{MET_PTcut}
\end{eqnarray}

We show the $\sigma\times {\rm BRs}$ of signal and backgrounds before and after cuts, as well as cut efficiencies (with respect to last level of cuts) for  $m_{H^0/A}=300$ GeV and $m_{H^\pm}=170$ GeV in Table~\ref{cuts1} for $\tau^\pm\to \pi^\pm\nu$.   We choose nominal value of 10 fb for the signal cross section\footnote{We use $\sigma(gg\rightarrow A/\H)=0.5$ pb, ${\rm BR}(A/\H\rightarrow W^\pm H^\mp)=100\%$, ${\rm BR}(H^\pm\rightarrow \tau \nu)=100\%$ to get the nominal signal cross section for $\tau \rightarrow \pi \nu$ final states. In general, the signal cross section depends on $\sin(\beta-\alpha)$ and $\tan\beta$ for a given value of $m_{H^0/A}$ and $m_{H^\pm}$. }.   The dominant background after cuts is the irreducible backgrounds $WW$ with one $W$ decaying leptonically and the other decaying to tau.  Utilizing the $\met$ cut and $p_T$ cut, all the backgrounds could be suppressed sufficiently.

\begin{figure}[h!]
\begin{center}
 \includegraphics[scale=1,width=7cm]{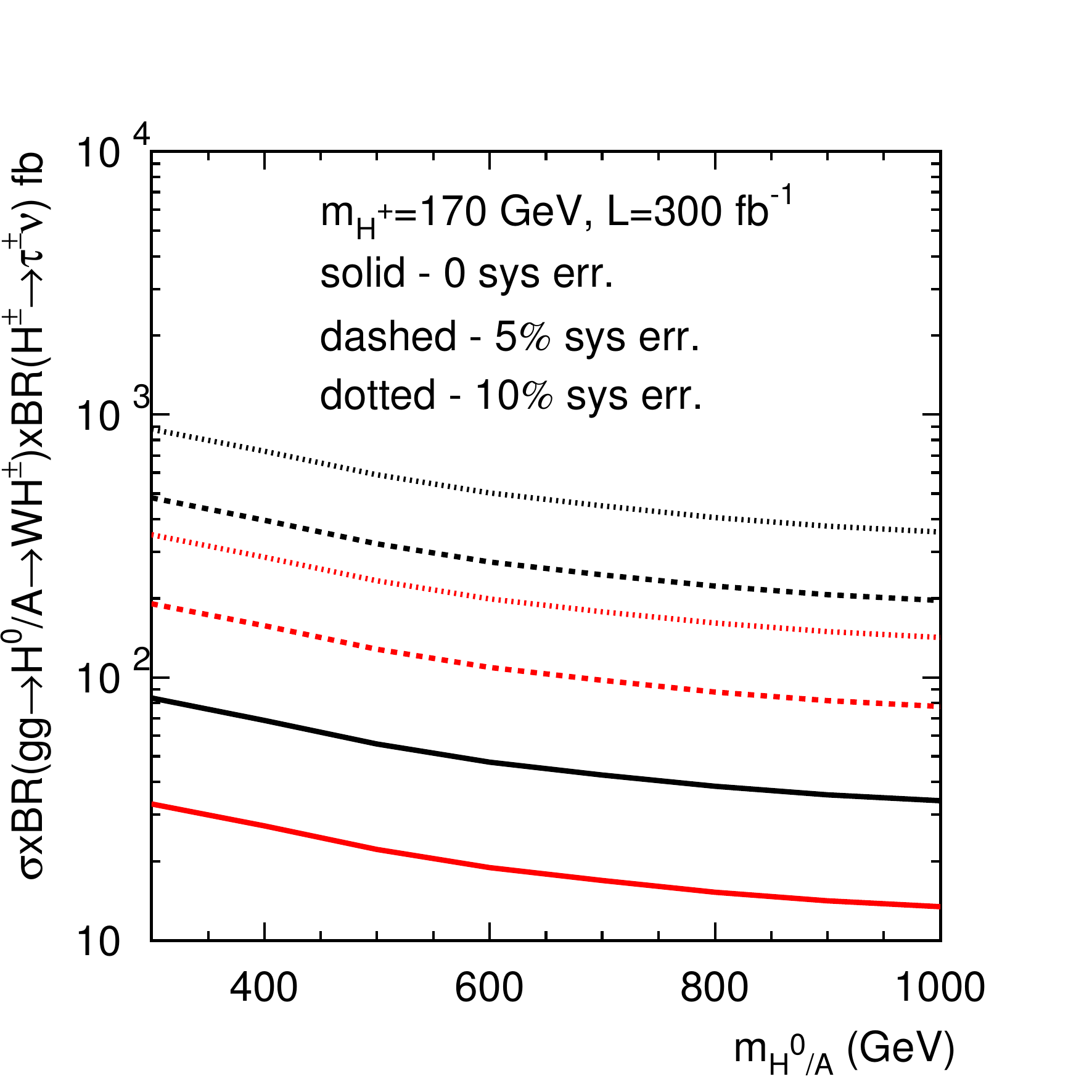}
\includegraphics[scale=1,width=7cm]{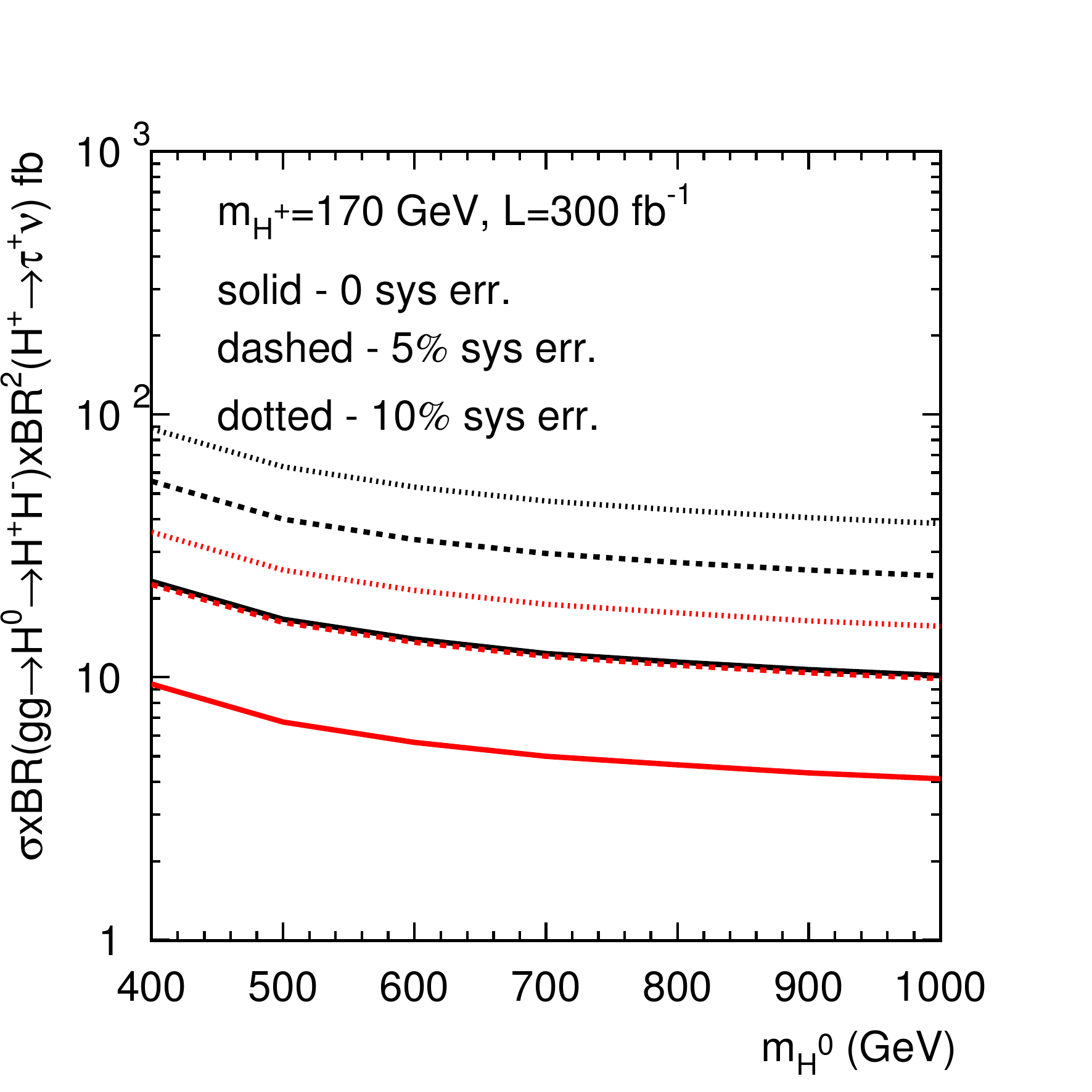}
 \end{center}
\caption{The 95\% CL exclusion limit (red lines)  and 5$\sigma$ discovery reach (black lines)  of $\sigma\times {\rm BR}(gg\to A/\H \to W^\mp H^\pm)\times {\rm BR}(H^+\to \tau^+\nu)$ vs. $m_{H^0/A}$ (left panel) and  $\sigma\times {\rm BR}(gg\to H^0\to H^+ H^-)\times {\rm BR}^2(H^+\to \tau^+\nu)$ vs. $m_{H^0}$ (right panel) at the 14 TeV LHC with 300 ${\rm fb}^{-1}$ luminosity, with $m_{H^\pm}$ fixed at 170 GeV.  The solid, dashed and dotted lines are for the limits with no systematic error, 5\% and 10\% systematic errors, respectively.
}
\label{limit1}
\end{figure}

The left panel of Fig.~\ref{limit1} shows the 95\% C.L. exclusion limit (red lines) and 5$\sigma$ discovery reach (black lines) for $\sigma\times {\rm BR}(gg\to A/\H\to W^\mp H^\pm)\times {\rm BR}(H^+\to \tau^+\nu)$ as a function of $m_{H^0/A}$ at the 14 TeV LHC with 300 ${\rm fb}^{-1}$ luminosity, combining both $\pi$ and $\rho$ channels.  We have fixed $m_{H^\pm}=170$ GeV.  The solid, dashed and dotted lines are for the limits with no systematic error, 5\% and 10\% systematic errors, respectively.  For the neutral Higgs mass between 300 to 1000 GeV, the cross section limits vary between 30 to 10 fb for 95\% C.L. exclusion and about 80 to 30 fb for 5$\sigma$ discovery.  The cross section limits with 5\% or 10\% systematic error included are about a factor of 7 or 10 worse.

\subsection{$gg \rightarrow H^0 \rightarrow H^+ H^-$}
\label{sec:HpmHmp}

Another interesting search channel with the charged Higgs in the final states is
\begin{eqnarray}
&&gg\to H^0\to H^+ H^-\to \tau^+ \tau^- \nu \bar{\nu},
\end{eqnarray}
with both taus decaying hadronically.
The leading SM backgrounds are
\begin{eqnarray}
W^+W^-\to \tau^+\tau^-  \nu_\tau\bar{\nu}_\tau,\  \ ZZ\to \tau^+\tau^-\nu\bar{\nu},\ \
W^\pm Z\to \ell^\pm \nu_\ell \tau^+\tau^-. 
\end{eqnarray}
The $W^\pm Z$ background can be reduced by lepton veto.

\begin{figure}[h!]
\begin{center}
\includegraphics[scale=1,width=7cm]{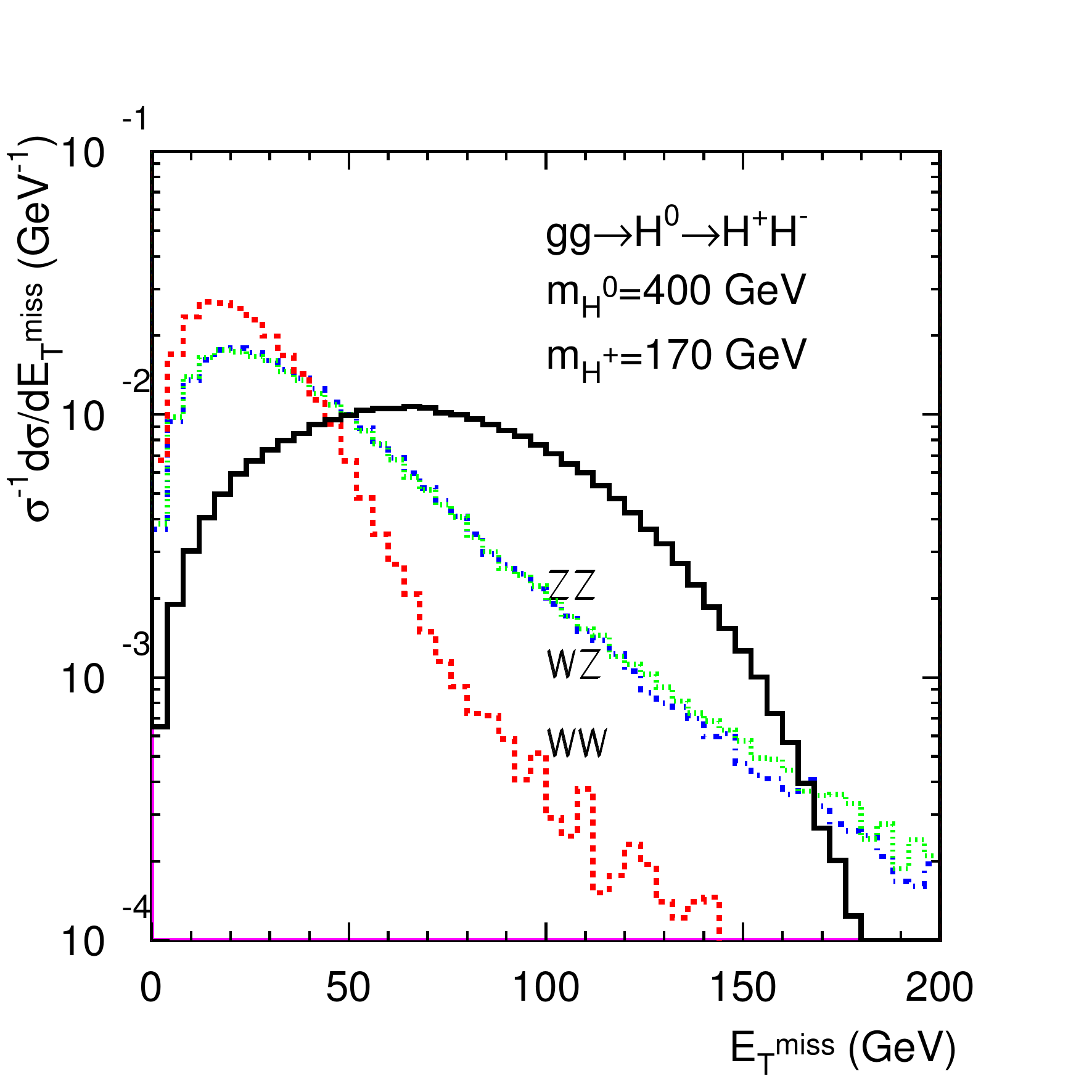}
 \includegraphics[scale=1,width=7cm]{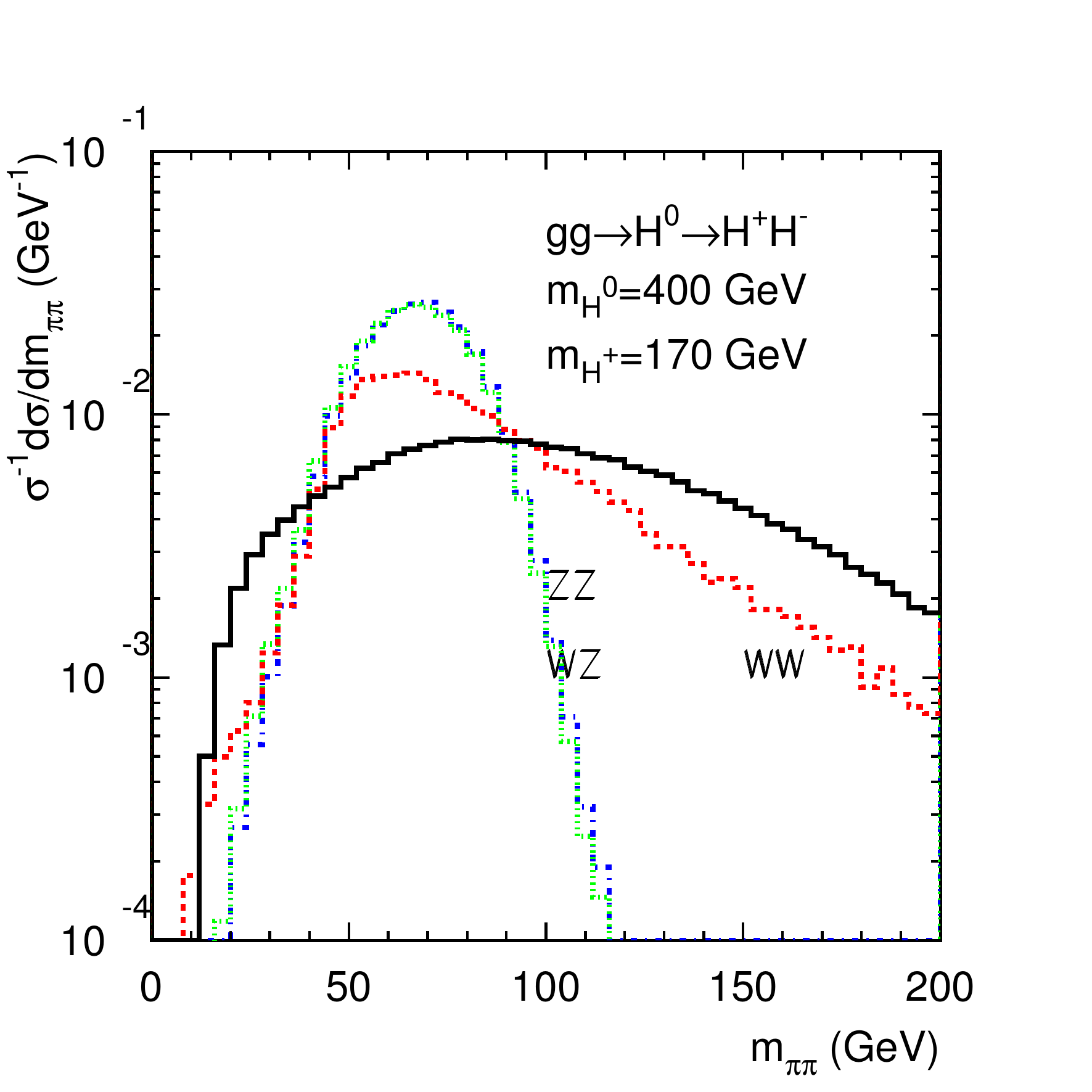}
\end{center}
\caption{The distributions of $\met$,  and $m_{\pi\pi}$ for the signal $H^+H^-$ (black curves) and the SM backgrounds after basic cuts at the 14 TeV LHC in $\tau\rightarrow \pi \nu$ final states. We assume $m_{H^0}=400$ GeV and $m_{H^\pm}=170$ GeV.   }
\label{events143}
\end{figure}

Adopting the same basic cuts as in Eqs.~(\ref{basic1}) and (\ref{veto1}),   the distributions of $\met$ and the invariant mass $m_{\pi \pi}$ for the signal and the SM backgrounds are shown in Fig.~\ref{events143}.   The $p_T(\pi)$ distribution is very similar to that of Fig.~\ref{events14}.  Both $\met$ and $p_T(\pi,\rho)$ peak at higher value, similar to the $W^\pm H^\mp$ channel.     In addition to the $p_T (\pi, \rho)$ and $\met$ cuts the same as in Eq.~(\ref{MET_PTcut}),  we impose $m_{\pi\pi, \rho\rho}$ cuts to reduce the $ZZ$ and $WZ$ backgrounds with $\tau\tau$ pair coming from $Z$ boson:
\begin{eqnarray}
  m_{\pi\pi,\rho\rho}< 50 \ {\rm GeV}\ {\rm or}\ m_{\pi\pi,\rho\rho}>90 \ {\rm GeV}.
\end{eqnarray}

\begin{table}[h!]
\begin{center}
\begin{tabular}{|c|c|c|c|c|c|c|}
  \hline
  Cuts & Signal & $WW$ & $ZZ$ & $WZ$ & $S/B$ & $S/\sqrt{B}$\\
  \hline
  $\sigma\times$BRs (fb) & 1 & 16 & 4.6 & 4.2 & 0.04 & 3.5\\
  \hline
 basic & 0.62 & 0.04 & 0.14 & 0.01 & - & - \\
 \hline
 $\met$ & 0.7 & 0.1 & 0.33 & 0.2 & - & -\\
 \hline
 $p_T^{\rm max}(\pi)$ & 0.88 & 0.86 & 0.77 & 0.75 & - & -\\
 \hline
 veto $m_{\pi\pi}$ & 0.72 & 0.87 & 0.18 & 0.17 & - & -\\
 \hline
 \hline
  $\sigma\times$BRs (fb) & 0.28 & 0.048 & 0.029 & 0.001 & 3.5 & 17 \\
 \hline
\end{tabular}
\end{center}
\caption{The cross sections and cut efficiencies  of the signal $gg\to H^0\to H^+H^- \rightarrow \tau^+ \tau^- \nu \bar{\nu}$ and the SM backgrounds after various cuts for  $\tau^\pm\to \pi^\pm\nu$ channel at the 14 TeV LHC. We assume $L=300$ fb$^{-1}$ and a nominal cross section for the signal to be 1 fb.  We fix $m_{H^0}=400$ GeV and $m_{H^\pm}=170$ GeV.  The $K$-factors for backgrounds are included.
}
\label{cuts3}
\end{table}

In Table~\ref{cuts3}, we show cut efficiencies as well as the signal and the SM background cross sections before and after cuts, for $gg\rightarrow H^0 \rightarrow H^+H^-$ with $H^\pm \rightarrow \tau \nu$ for $\tau^\pm\to \pi^\pm\nu$ final states.
The dominant SM background is $WW$, which can be suppressed sufficiently to achieve good signal significance.

In the right panel of Fig.~\ref{limit1}, the 95\% C.L. exclusion limit (red lines) and 5$\sigma$ discovery reach (black lines) for $\sigma\times {\rm BR}(gg\to H^0\to H^+ H^-)\times {\rm BR}^2(H^+\to \tau^+\nu)$  are shown as a function of $m_{H^0}$ at the 14 TeV LHC with 300 ${\rm fb}^{-1}$ luminosity, combining both $\pi$ and $\rho$ channels.   We have fixed $m_{H^\pm}=170$ GeV.  For the neutral Higgs mass between 400 to 1000 GeV, the cross section limits vary between 9 to 4 fb for 95\% C.L. exclusion and about 20 to 10 fb for 5$\sigma$ discovery.  The cross section limits with 5\% or 10\% systematic error included are about a factor of 2 or 4 worse.

\section{Implication for the Type II 2HDM}
\label{sec:implication}

 \begin{table}[h!]
\begin{center}
  \begin{tabular}{|l|c|c|c|c| }
    \hline
    $\left\{ {m_{\H},m_{\A}}\right\}$ GeV  & $\A\to W^\pm H^\mp$ & $\H\to W^\pm H^\mp $ &$\H\to H^+H^-$  &$\H\to \A Z $     \\ \hline
     BP1: $\left\{ {1000, 300}\right\}$ & \cmark &  $-$ &$-$ &$-$   \\ \hline
    BP2: $\left\{ {300, 1000}\right\}$ &$-$ & \cmark  & \xmark& \xmark \\ \hline
   BP3: $\left\{ {400, 1000}\right\}$  & $-$& \cmark &\cmark & \xmark   \\ \hline
     \end{tabular}
\end{center}
\caption{Benchmark points shown for illustrating the discovery and exclusion limits of $gg\rightarrow A /\H \rightarrow W^\pm H^\mp$ and $gg \rightarrow \H \rightarrow H^+H^-$ in the context of the Type II 2HDM. We assume $m_{H^\pm}=170$ GeV and $m_{\h}=126$ GeV. The checkmarks indicate kinematically allowed channels.  The ``$-$'' means the parent particle being too heavy to be of interesting.   }
\label{table:bmp}
\end{table}

To interpret the 95\% C.L. exclusion and 5$\sigma$ reach limits in the type II 2HDM, we choose three benchmark points as listed in Table.~\ref{table:bmp}, with $m_{H^\pm}$ fixed to be 170 GeV and $\h$ being the observed 126 GeV SM-like Higgs.  BP1 and BP2 with $(m_{H^0},m_A)$=(1000, 300) GeV or (300, 1000) GeV are the best case scenario for  $A\rightarrow W^\pm H^\mp$ and  $H^0\rightarrow W^\pm H^\mp$ as   other competing decays of $A$ and $H^0$ are kinematically unaccessible.  BP3 with $(m_{H^0},m_A)$=(400, 1000) GeV is chosen to illustrate the reach of $H^0 \rightarrow H^+H^-$ with a suppressed decay branching fraction of $H^0 \rightarrow W^\pm H^\mp$.

\begin{figure}[h!]
\begin{center}
\includegraphics[scale=1,width=7cm]{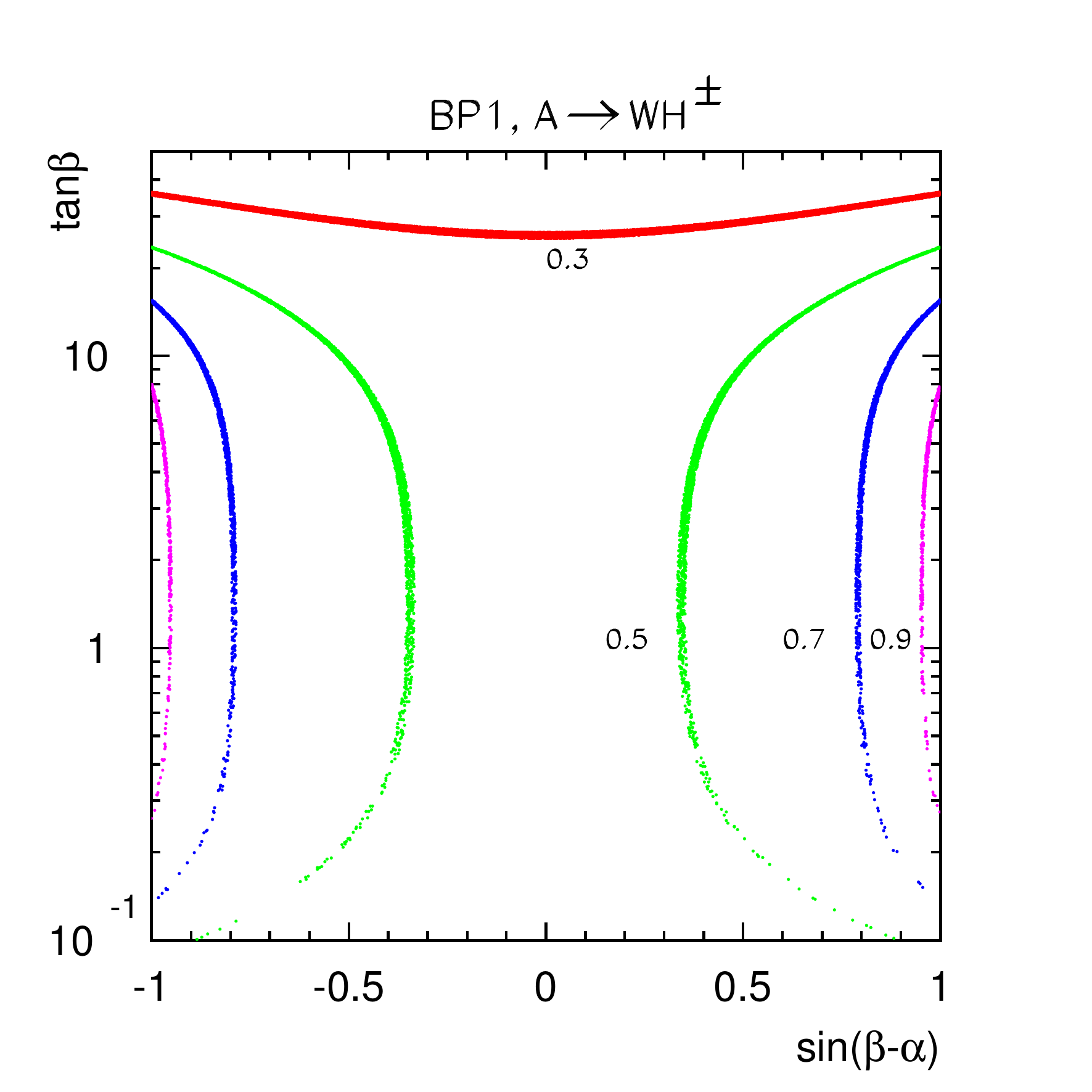}
\includegraphics[scale=1,width=7cm]{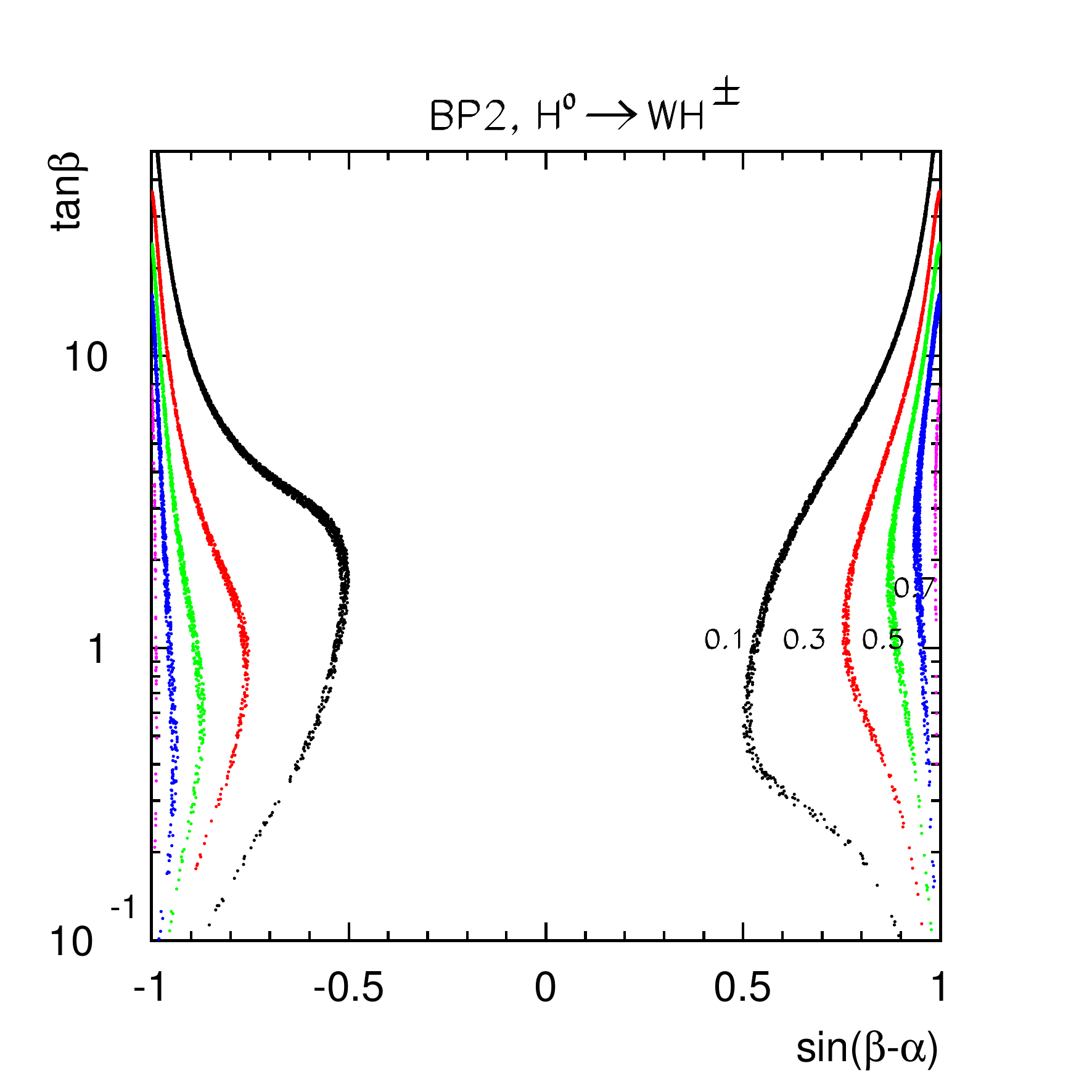}
 \end{center}
\caption{Branching fractions of $A \rightarrow W^\pm H^\mp$  for BP1 (left panel) and $\H \rightarrow W^\pm H^\mp$ for BP2 (right panel), respectively.  We assume $m_{H^\pm}=170$ GeV and $m_{h^0}=126$ GeV.   }
\label{Fig:contour_BP12}
\end{figure}

The production of $gg \rightarrow A$ only depends on $\tan\beta$ while $gg \rightarrow H^0$ depends on both $\tan\beta$ and $\sin(\beta-\alpha)$~\cite{Coleppa:2014hxa}.   In Fig.~\ref{Fig:contour_BP12}, the branching fractions of $A/H^0 \rightarrow W^\pm H^\mp$ are shown for BP1 (left panel) and BP2 (right panel), respectively.   The suppression of the ${\rm BR}(A\rightarrow W^\pm H^\mp)$ at $\sin(\beta-\alpha)\sim 0$ is due to the competing $A \rightarrow h^0 Z$ mode.    ${\rm BR}(A\rightarrow W^\pm H^\mp)$ gets smaller at both large and small  $\tan\beta$ due to the competing $A$ decays to fermions or loop induced processes.   For $H^0\rightarrow W^\pm H^\mp$,  the branching fraction is larger than 50\% only for $|\sin(\beta-\alpha)| \gtrsim 0.8$, since $H^0W^\pm H^\mp$ coupling is proportional to $\sin(\beta-\alpha)$.

\begin{figure}[h!]
\begin{center}
\includegraphics[scale=1,width=7cm]{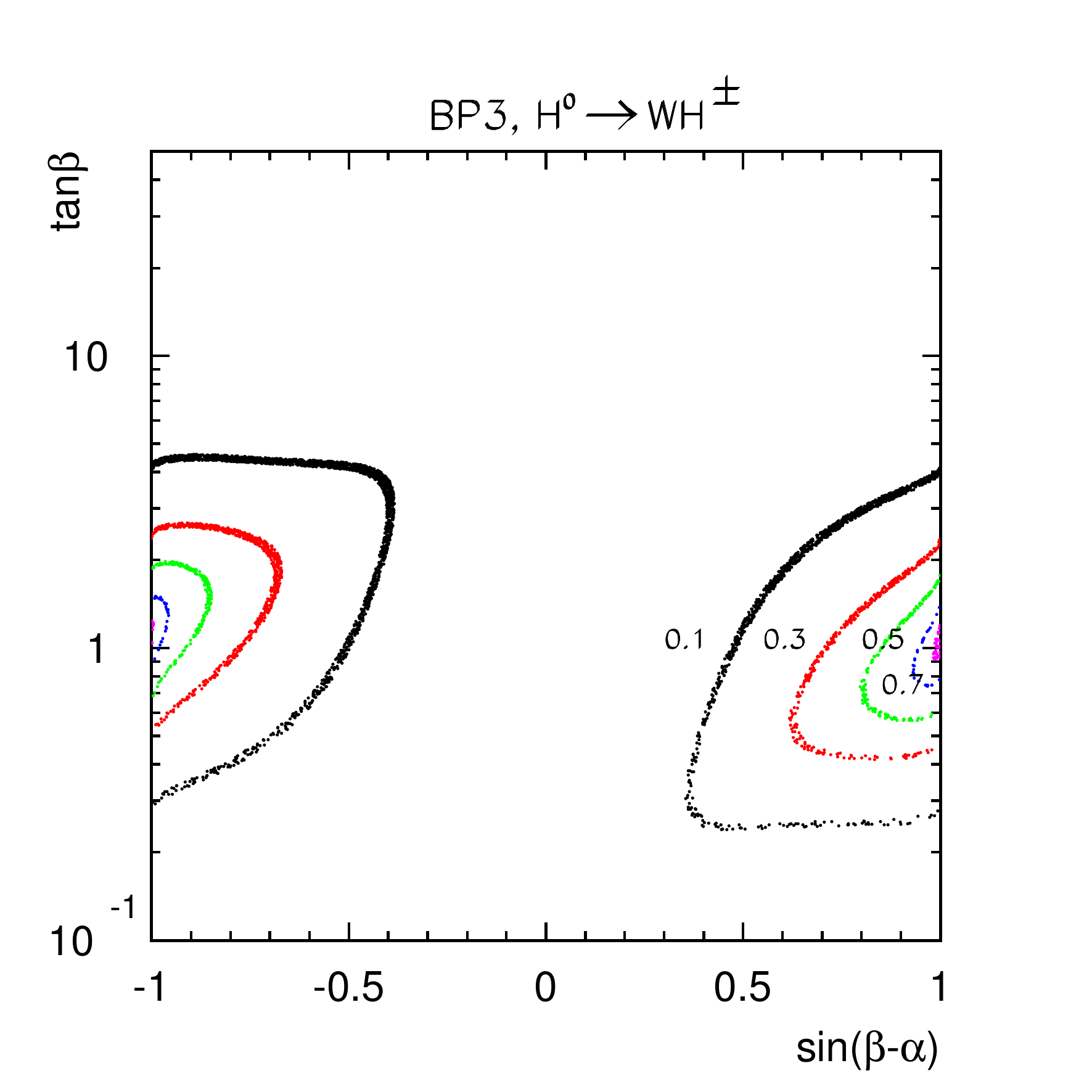}
\includegraphics[scale=1,width=7cm]{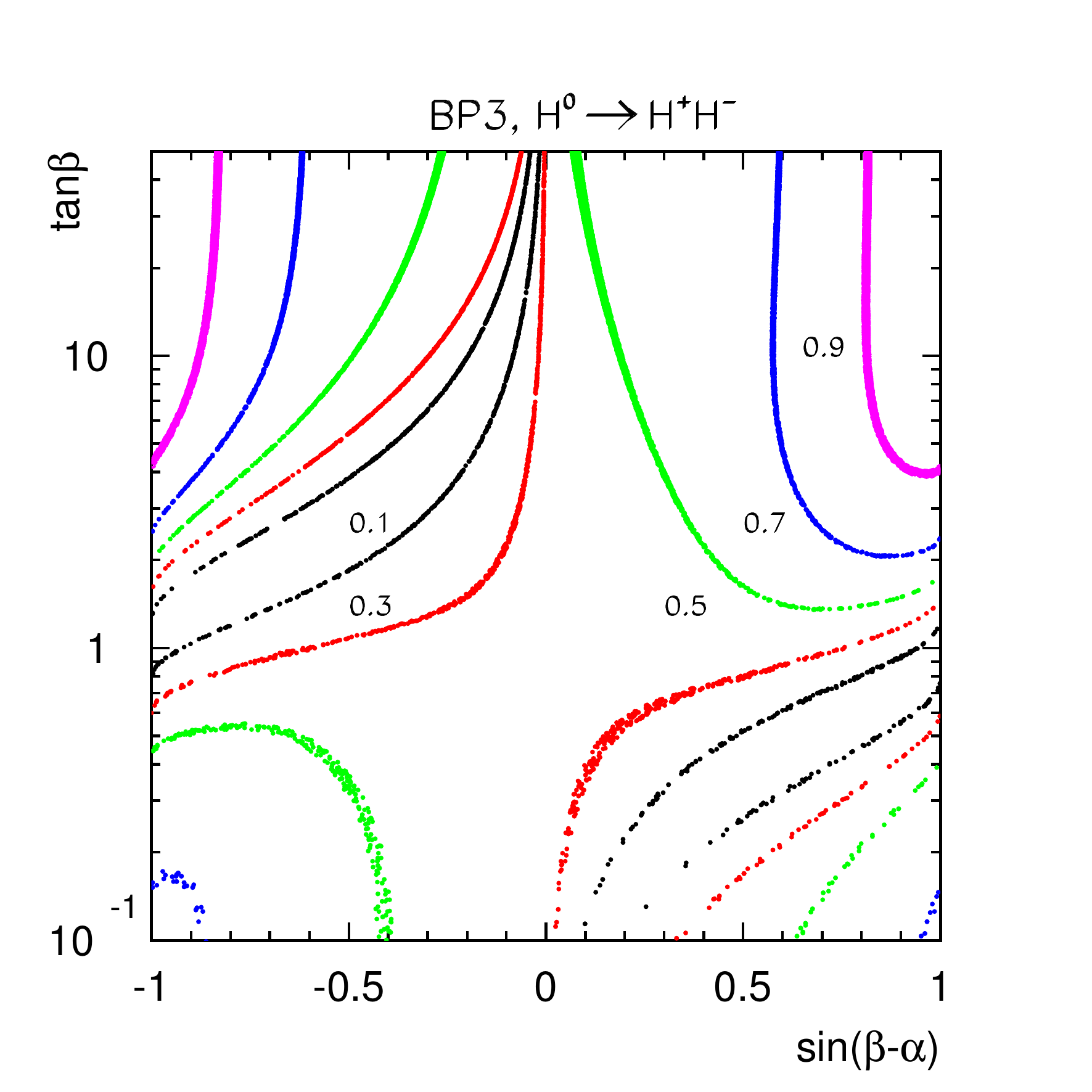}
 \end{center}
\caption{Branching fractions of $H^0 \rightarrow W^\pm H^\mp$ (left panel)  and $H^0 \rightarrow H^+ H^-$  for BP3, respectively.  We assume $m_{H^\pm}=170$ GeV, $m_{h^0}=126$ GeV and $m_{12}^2=0$. }
\label{Fig:contour_BP3}
\end{figure}

In Fig.~\ref{Fig:contour_BP3},  the branching fractions of $H^0 \rightarrow W^\pm H^\mp$  and $H^0 \rightarrow H^+ H^-$  are shown in the left and right panel, respectively, for BP3.  Comparing to BP2 with $m_{H^0}=300$ GeV,  ${\rm BR}(H^0 \rightarrow W^\pm H^\mp)$ receives stronger suppression at both small and large $\tan\beta$ due to the opened $H^0\rightarrow H^+H^-$ and $H^0\rightarrow t\bar{t}$.   It is only significant around a small region near $\tan\beta \sim 1$.

The branching fraction of $H^0 \rightarrow H^+ H^-$ exhibits more complicated dependence on $\sin(\beta-\alpha)$ and $\tan\beta$ due to the $H^0H^+H^-$ coupling.     The branching fraction is more than 50\% at $\tan\beta\gtrsim 2$ or $\tan\beta\lesssim 0.5$  for $|\sin(\beta-\alpha)|$ close to 1.

\begin{figure}[h!]
\begin{center}
\includegraphics[scale=1,width=7cm]{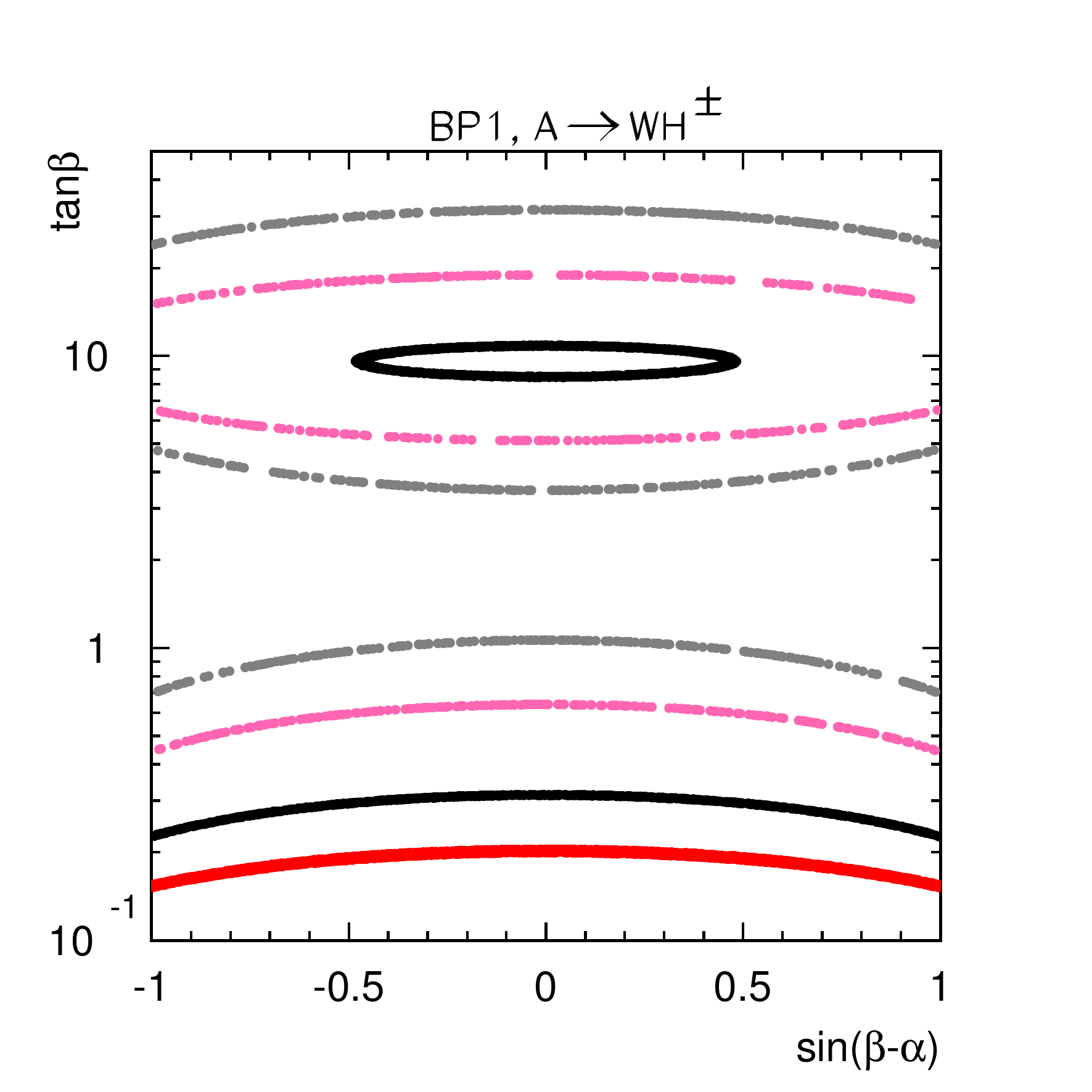}
\includegraphics[scale=1,width=7cm]{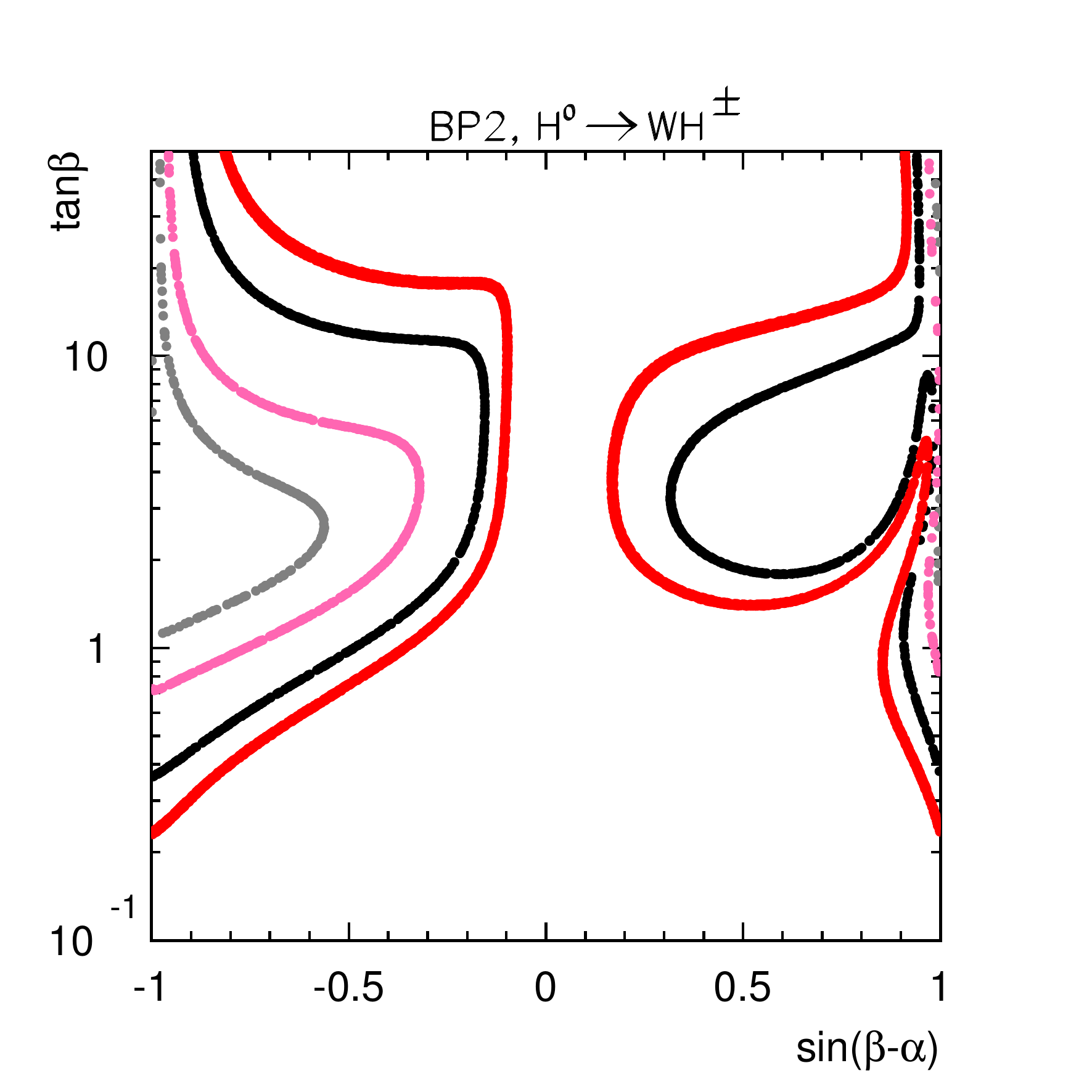}\\
 \end{center}
\caption{Reach for $gg\to H^0/A\to W^\mp H^\pm$  for BP1 (left panel) and BP2 (right panel) at the 14 TeV in $\sin(\beta-\alpha)$ versus $\tan\beta$ plane in the type II 2HDM at the 14 TeV LHC with 300 ${\rm fb}^{-1}$ luminosity. The red (pink)  and black(grey) lines are $95\%$ CL exclusion and $5\sigma$ discovery reach with 0 (10\%) systematic error.   See text for details.}
\label{Fig:contour1}
\end{figure}

In the left panel of Fig.~\ref{Fig:contour1}, we show the reach in $\sin(\beta-\alpha)$ versus $\tan\beta$ plane for $A\rightarrow W^\pm H^\mp$ (BP1).    For BP1 with $m_A=300$ GeV,  regions above the red (black) curves can be excluded at 95\% C.L. (discovered at 5$\sigma$) at the 14 TeV with 300 ${\rm fb}^{-1}$ luminosity, except the region enclosed by the black ellipse around $\tan\beta\sim 10$, which can not be covered by 5$\sigma$ discovery due to the suppression of the production cross section.  The region with $\tan\beta\gtrsim 0.2$ can be covered by exclusion for all values of $\sin(\beta-\alpha)$ and $\tan\beta\gtrsim 0.3$ can be covered by discovery except the region with $\tan\beta\sim 10, |\sin(\beta-\alpha)|<0.5$.   The loss of sensitivity at small $\tan\beta$ is mainly due to the reduction of $H^\pm \rightarrow \tau\nu$ branching fraction.    Reach gets slightly worse for $\sin(\beta-\alpha)$ close to zero due to the competition of $A\rightarrow Zh^0$, which suppresses the branching fraction of $A\rightarrow W^\pm H^\mp$ correspondingly (See the left panel of Fig.~\ref{Fig:contour_BP12}).   Reach with 10\% systematic error are shown in pink and grey curves for 95\% C.L. exclusion and 5$\sigma$ discovery, respectively.   The regions shrink considerably comparing to the 0 systematic error case.  Only $0.6 < \tan\beta < 5$ and $\tan\beta > 20$ can be excluded and the discovery reach is further reduced to   $1 < \tan\beta < 3$ or $\tan\beta>30$.

In the right panel of Fig.~\ref{Fig:contour1}, we show the reach in $\sin(\beta-\alpha)$ versus $\tan\beta$ plane for $H^0\rightarrow W^\pm H^\mp$ (BP2), indicated by  regions to the left (right) of the curves for negative (positive)  $\sin(\beta-\alpha)$.   Regions with $\sin(\beta-\alpha)<-0.2$ or   $\sin(\beta-\alpha)>0.3$  can be covered by $gg\rightarrow H^0 \rightarrow W^\pm H^\mp$ with 95\% C.L. exclusion.  This channel is insensible to region around $\sin(\beta-\alpha) \sim 0$ since $H^0 \rightarrow W^\pm H^\mp$ is highly suppressed.   This channel is sensitive to intermediate $\tan\beta$ between 1 to 10, while the reach for $\tan\beta$ is enhanced for $|\sin(\beta-\alpha)| \sim 1$, which is the preferred region for $h^0$ being the SM-like 126 GeV Higgs.   The decrease in the sensitivity in the thin slice region for $\sin(\beta-\alpha)$ between 0.6 to 0.9 is due to the reduction of $gg\rightarrow H^0$ production cross section.  The reach for 5$\sigma$ discovery is slightly worse.  Introducing 10\% systematic error (regions enclosed by the pink and grey curves)   reduces the exclusion and discovery reach further.

The reach of $H^0\rightarrow W^\pm H^\mp$ for BP3 with $m_{H^0}=400$ GeV is similar to that of BP2.  There is no reach in regions of $\tan\beta\gtrsim 10$ and $\tan\beta\lesssim 0.4$, due to the competing $H^0\rightarrow H^+H^-, t\bar{t}$ modes.

\begin{figure}[h!]
\begin{center}
  \includegraphics[scale=1,width=7cm]{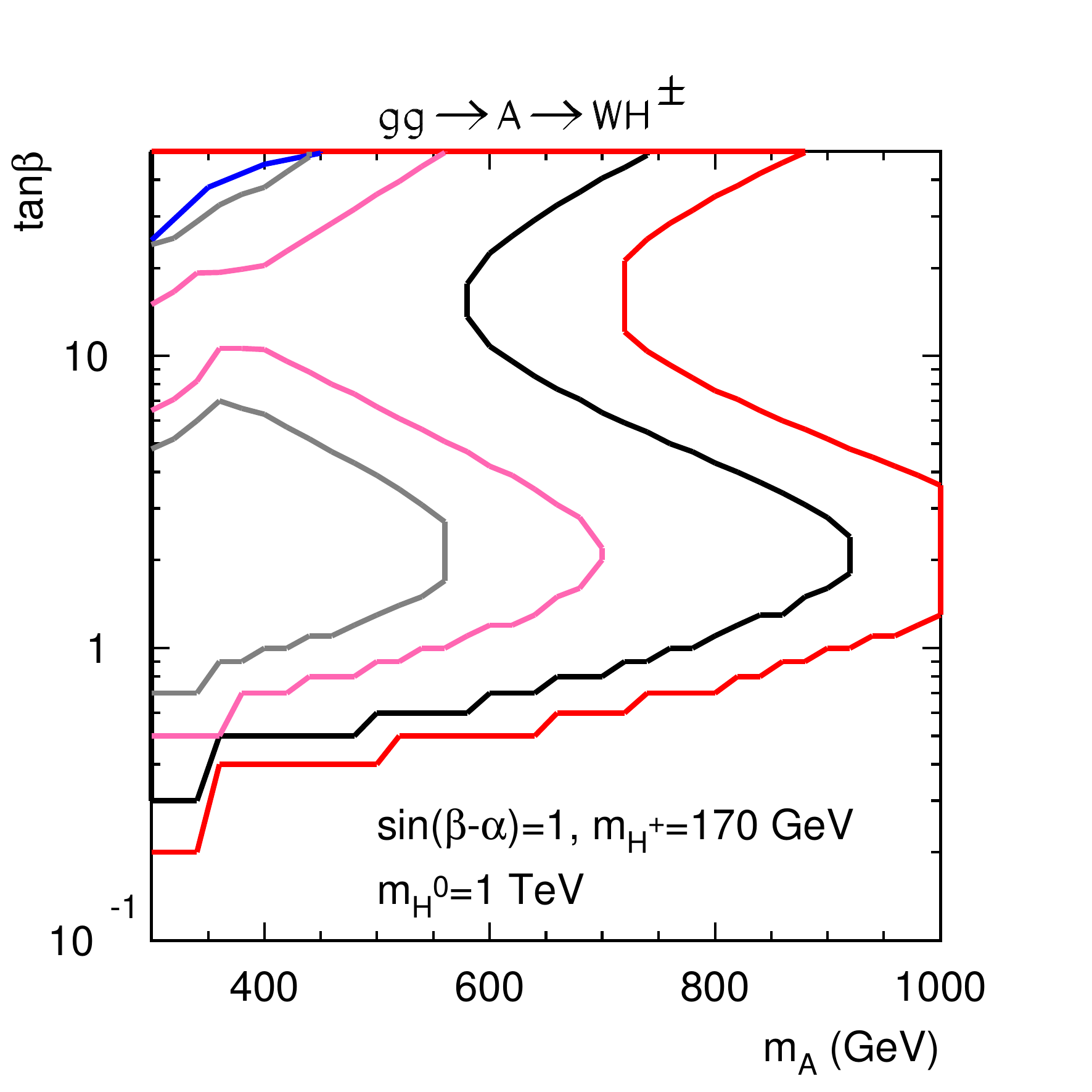}
\includegraphics[scale=1,width=7cm]{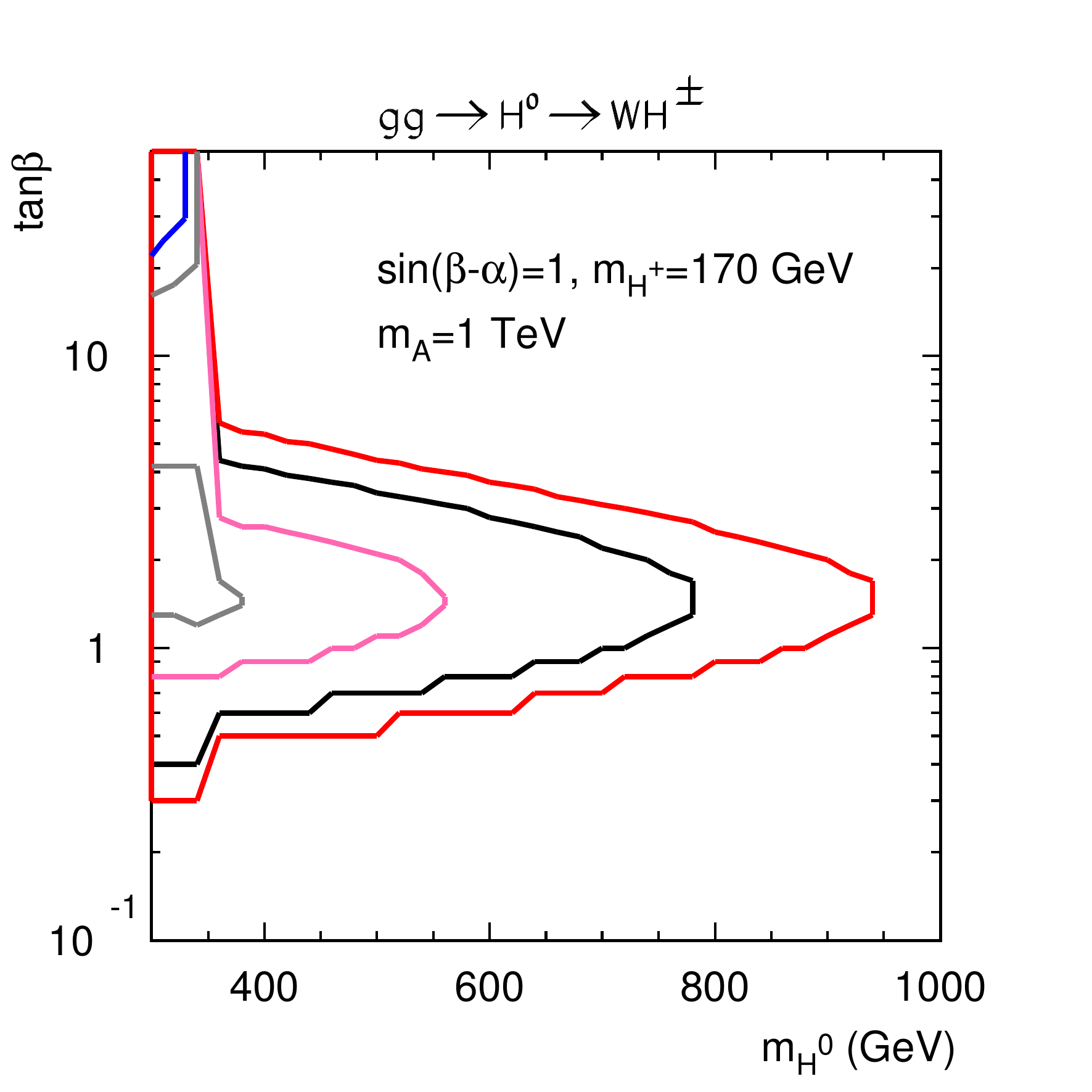}
\end{center}
\caption{Reach for $gg\to A/H^0\to W^\pm H^\mp$ at the 14 TeV with 300 ${\rm fb}^{-1}$ luminosity  in $m_A$ versus $\tan\beta$ plane (left panel),  $m_{H^0}$ versus $\tan\beta$ plane (right panel) in the type II 2HDM.   The blue curve shows the reduced exclusion by $H^0/A\to \tau^+\tau^-$~\cite{Aad:2014vgg} due to the opening of $A/\H  \rightarrow W^\pm H^\mp$. Regions to the left of the red/pink (black/grey) curve can be excluded (discovered) assuming  0 and 10\% systematic errors, respectively. We assume $m_{H^\pm}=170$ GeV and $\sin(\beta-\alpha)=1$.
}
\label{Fig:MHTB_WHpm}
\end{figure}

Fig.~\ref{Fig:MHTB_WHpm} shows the exclusion (region to the left of red/pink curve) and discovery (region to the left of black/grey curve) reach in $m_{A/H^0}$ versus $\tan\beta$ plane for $A$ (left panel) and $H^0$ (right panel) with $gg\rightarrow A/H^0 \rightarrow W^\pm H^\mp$ channel.    For low $m_A$ of 300 GeV, $\tan\beta$ as low as 0.2 can be probed, while $\tan\beta>1$ can be excluded for $m_A=1000$ GeV.    Reach is reduced for $\tan\beta$ around 10 due to the suppression of the production cross section.  Also shown in blue curve is the reduced exclusion by $H^0/A\to \tau^+\tau^-$~\cite{Aad:2014vgg} due to the opening of $A \rightarrow W^\pm H^\mp$.    The reach for the exotic decay of $A \rightarrow W^\pm H^\mp$ can cover most of the parameter region in $m_A$ versus $\tan\beta$ plane while the conventional search mode of $A \rightarrow \tau\tau$ mode being highly suppressed.  For results with 10\% systematic error included (regions to the left of pink/grey curves), the mass reach is about 350 GeV less.

The reach for $H^0 \rightarrow W^\pm H^\mp$ is much more limited, in particular for $m_{H^0}>400$ GeV since $H^0\rightarrow W^\pm H^\mp$ is suppressed once $H^0 \rightarrow H^+H^-$ opens for $m_{H^0}>2m_{H^\pm}$.   Regions around $\tan\beta\sim 1$, however, can still be excluded (discovered) for $m_{H^0}$ up to 900 (800) GeV due to the suppression of $H^0\rightarrow H^+H^-$ in that region.   Given the insensitivity of conventional search channel $H^0\rightarrow \tau\tau$ for $\tan\beta\sim 1$, $H^0\rightarrow W^\pm H^\mp$ could be an alternative discovery channel for the neutral CP-even Higgs.   For results with 10\% systematic error included, the reach is about 400 GeV less.

\begin{figure}[h!]
\begin{center}
\includegraphics[scale=1,width=7cm]{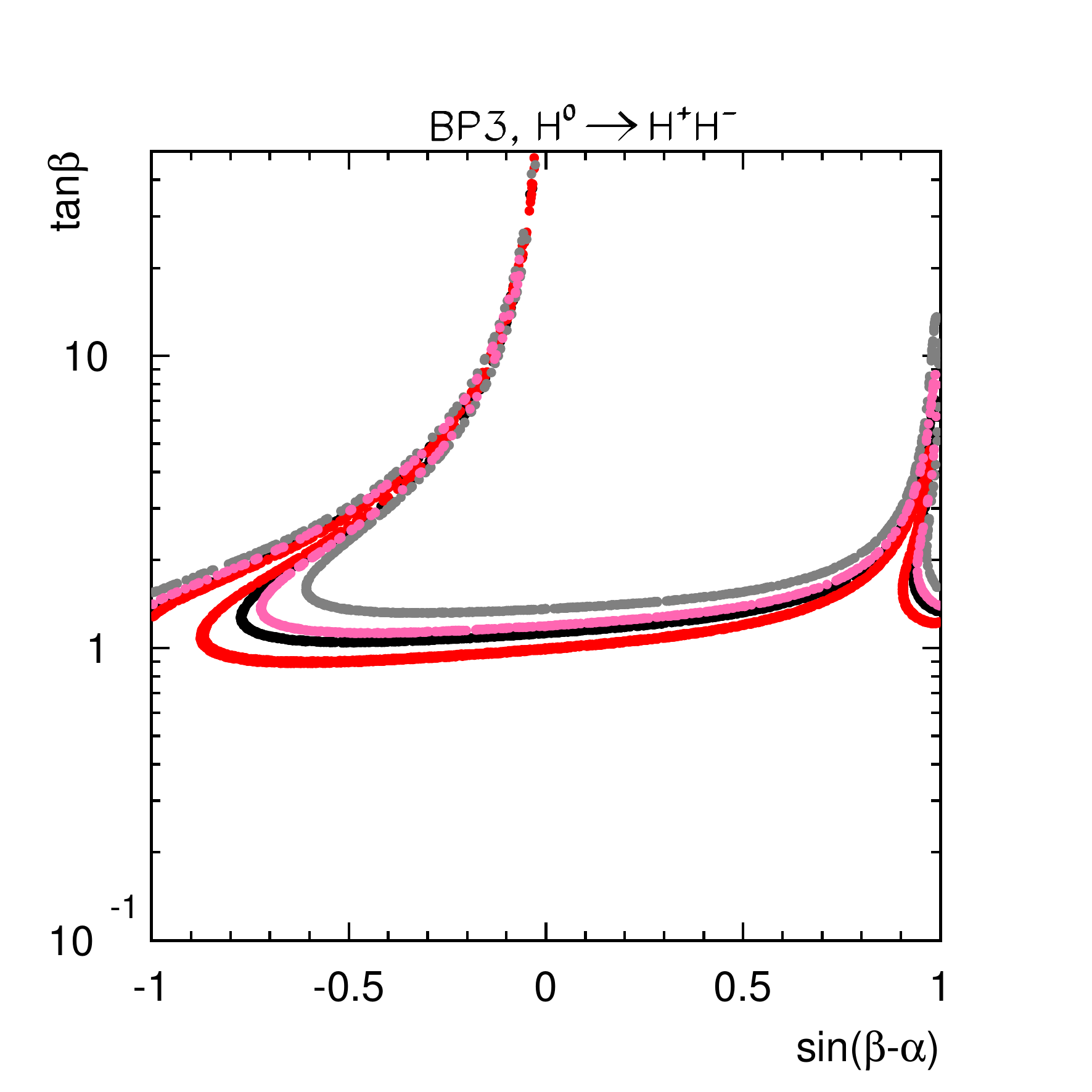}
 \includegraphics[scale=1,width=7cm]{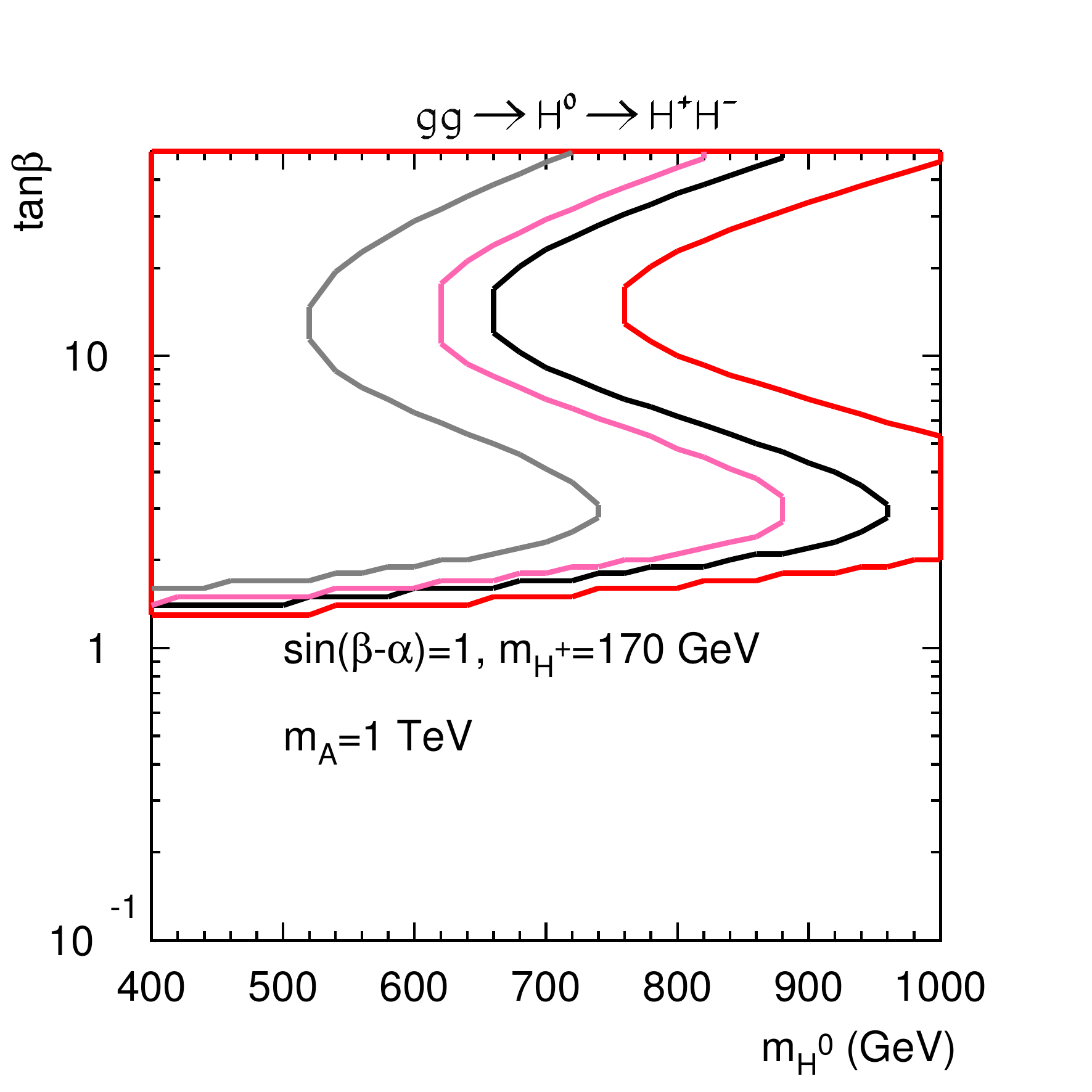}
\end{center}
\caption{Reach for $gg\to H^0\to H^+ H^-$ at the 14 TeV with 300 ${\rm fb}^{-1}$ luminosity in $\sin(\beta-\alpha)$ versus $\tan\beta$ plane for BP3 with $m_{H^0}=40$0 GeV (left panel) and $m_{H^0}$ versus $\tan\beta$ plane with $\sin(\beta-\alpha)=1$ (right panel) in the type II 2HDM.   Regions above (to the left of) the red/pink and black/grey lines are $95\%$ CL exclusion and $5\sigma$ discovery reach for the left (right)   panel, assuming 0 and 10\% systematic errors, respectively. We assume $m_{H^\pm}=170$ GeV.    }
\label{Fig:contour2}
\end{figure}

In the left panel of Fig.~\ref{Fig:contour2}, we show the 95\% C.L. exclusion (red/pink)  and 5$\sigma$ discovery (black/grey) reach (regions   above the curves) in $\sin(\beta-\alpha)$ versus $\tan\beta$ plane for $H^0\rightarrow H^+ H^-$ for BP3.    This channel is sensitive to large region of the parameter space for $\tan\beta>1$.  The thin slice of insensitive region at  negative   $\sin(\beta-\alpha)$ is due to  the suppression of $H^0\rightarrow H^+H^-$,  while the  thin slice of insensitive region at  positive   $\sin(\beta-\alpha)$ is due to the suppression of $gg \rightarrow H^0$.

 The  right panel of Fig.~\ref{Fig:contour2} shows the reach in $m_{H^0}$ versus $\tan\beta$ plane  with $gg\rightarrow H^0\rightarrow H^+ H^-$ channel.  Regions with $\tan\beta \gtrsim 1-2$ can be excluded at 95\% C.L. for $m_{H^0}  $ between 400 to 1000 GeV.  For results with 10\% systematic error included, the mass reach is about 150 GeV less.   The reach around $\tan\beta \sim10$ is reduced due to the reduction in the production cross section.    Combing both $H^0\rightarrow W^\pm H^\mp$ and $H^0 \rightarrow H^+H^-$ channels,  most regions of $\tan\beta \gtrsim 0.3$ can be covered, while the conventional search channel of $H^0\rightarrow \tau\tau$ gets highly suppressed.

\section{Conclusion}
\label{sec:conclusions}

 The conventional search mode for the extra neutral Higgses in models with an extension of the SM Higgs sector is $A/H^0 \rightarrow \tau\tau$, which is only sensitive to the large $\tan\beta$ region.  Furthermore, the opening of the exotic Higgs decay modes, for example $A/H^0\rightarrow W^\pm H^\mp$ and $H^0\rightarrow H^+H^-$, greatly reduces the sensitivity of the $\tau\tau$ mode.   These new decay channels, however, can be used to probe the heavy neutral Higgses, which is discussed in detail in this paper.

One feature of the taus from light $H^\pm$ decay is that the decay products of tau, for example, pions and rhos, tend to be more energetic comparing to the SM background taus from $W$ decay.  This can be used to suppress the SM backgrounds in both channels that we analysed:  $gg \rightarrow A/H^0 \rightarrow W^\pm H^\mp \rightarrow \ell \tau +\met$ and $gg \rightarrow H^0 \rightarrow H^+ H^- \rightarrow \tau \tau +\met$.  For $m_{A/H^0}$ between 300 and 1000 GeV, we found that the $\sigma\times{\rm BR}(gg \rightarrow A/H^0 \rightarrow W^\pm H^\mp)\times {\rm BR}(H^\pm \rightarrow \tau \nu)$ varies from 30 fb to 10 fb for 95\% exclusion, and about 80 to 30 fb for 5$\sigma$ discovery.  For $H^+H^-$ mode, 95\% C.L. limits on $\sigma\times {\rm BR}(gg\to H^0\to H^+ H^-)\times {\rm BR}^2(H^+\to \tau^+\nu)$ vary from 9 to 4 fb for $m_{H^0}$ between 400 and 1000 GeV, while the 5$\sigma$ reach is from 20 to 10 fb.   The cross section limits including 5\% or 10\% systematic error is considerably worse.

We further interpret the cross section limits in the Type II 2HDM parameter space.  For $A\rightarrow W^\pm H^\mp$, we found that almost all regions of the parameter space in $\sin(\beta-\alpha)$ versus $\tan\beta$ plane can be covered, except for very small $\tan\beta$ for a benchmark point of $m_A=300$ GeV.    It is also sensitive to most regions in $m_A$ versus $\tan\beta$ plane with $m_A$ up to about 1000 GeV, and $\tan\beta$ as low as 1.   It provides an alternative channel to search for the CP-odd Higgs when the conventional mode of $A \rightarrow \tau\tau$ becomes ineffective.

For the CP-even Higgs $H^0$, the most sensitive channel is $H^0\rightarrow H^+H^-$, which covers all regions of the parameter space in $\sin(\beta-\alpha)$ versus $\tan\beta$ plane except for $\tan\beta <1$ for a benchmark point of $m_{H^0}=400 $ GeV.   In $m_{H^0}$ versus $\tan\beta$ plane,  $m_{H^0}$ up to about 1 TeV can be excluded at 95\% C.L., while  $m_{H^0}$ up to about 950 GeV can be discovered at 5$\sigma$ significant level.   While the reach in $H^0\rightarrow W^\pm H^\mp$ is more limited, especially for $m_{H^0}>2 m_{H^\pm}$, it is complementary to $H^0\rightarrow H^+ H^-$ mode for regions of $\tan\beta \lesssim 1$.  If 10\% systematic error is assumed, the mass reach is typically about 150 $-$ 400 GeV less.

The discovery of extra Higgses besides the SM-like one would certainly be an unambiguous evidence for new physics beyond the SM.  The exotic decay modes of heavy Higgs decaying to two light Higgses, or one Higgs with one SM gauge boson provide alternative search channels, which could greatly enhance the discovery potential for heavy Higgses at current and future colliders.      Once those non-SM Higgses are discovered, kinematic reconstruction would provide important information about their mass spectrum.  A cross check with the indirect flavor and precision constraints would be complementary and lead to new hints towards new physics beyond the SM.

\acknowledgments
 We thank Felix Kling for helpful discussions on the flavor and precision constraints.    The work of T.L. is supported by the ARC Centre of Excellence for Particle Physics at the Terascale. S.S. was supported by  the Department of Energy under  Grant~DE-FG02-13ER41976.

\bibliographystyle{JHEP}

\end{document}